\documentstyle[epsfig]{ioplppt}
\newcommand{\pyth}{{\sc PYTHIA}}
\newcommand{\ari}{{\sc ARIADNE}}
\newcommand{\hwg}{{\sc HERWIG}}
\newcommand{\ftg}{{\sc F2GEN}}
\newcommand{\pho}{{\sc PHOJET}}
\newcommand{\etout}{\mbox{$E_{\rm t,out}$}}
\newcommand{\dkt}{\mbox{$dk_T^2/(k_T^2+k_0^2)$}}
\newcommand{\phiSep}{\mbox{$\phi_{\mathrm{sep}}\,$}}
\newcommand{\FTWOGAM}{\mbox{$F^{\gamma}_{2}$}}
\newcommand{\ltap}{\;\raisebox{-.4ex}{\rlap{$\sim$}} \raisebox{.4ex}{$<$}\;}

\renewcommand{\vec}[1]{\ifmmode
\mathchoice{\mbox{\boldmath$\displaystyle\bf#1$}}
{\mbox{\boldmath$\textstyle\bf#1$}}
{\mbox{\boldmath$\scriptstyle\bf#1$}}
{\mbox{\boldmath$\scriptscriptstyle\bf#1$}}\else
{\mbox{\boldmath$\bf#1$}}\fi}
\newcommand\kev{\mbox{keV}}
\newcommand\mev{\mbox{MeV}}
\newcommand\gev{\mbox{GeV}}
\newcommand\dedx{\mbox{d{\it E}/d{\it x}}}
\newcommand\CL{\mbox{C.L.}}
\begin{document}
{\flushright RAL-97-042\\hep-ph/9708478\\}
\vspace{-4mm}
\title{Two photon physics at LEP2}[Two photon physics]

\author{Susan Cartwright\dag\ and
Michael H Seymour\ddag\ (conveners),\\
Klaus Affholderbach\S,
Frank Close\ddag,
Glen Cowan\S,
Alex Finch\P,\\
Jan Lauber+,
Mark Lehto\dag\ and
Alison Wright\ddag}

\address{\dag University of Sheffield Department of Physics, Sheffield,
  S3~7RH. U.K.}
\address{\ddag Rutherford Appleton Laboratory, Chilton,
  Didcot, Oxfordshire, OX11~0QX. U.K.}
\address{\S Universit\"at Siegen, Fachbereich Physik, D-57068 Siegen, Germany}
\address{\P University of Lancaster, Lancaster, LA1~4YB. U.K.}
\address{+University College London, Gower Street, London, WC1E~6BT. U.K.}

\begin{abstract}
The working group on two photon physics concentrated on three main
subtopics: modelling the hadronic final state of deep inelastic
scattering on a photon; unfolding the deep inelastic scattering data to
obtain the photon structure function; and resonant production of
exclusive final states, particularly of glueball candidates.  In all
three areas, new results were presented.
\end{abstract}


\section{Introduction}

Two photon physics at LEP2 is in a somewhat different position from the
other topics covered in this workshop, in that the increase in beam energy
makes little difference to the physics.  There is a small increase in
cross-section ($\propto\ln E$), but the main effect is a considerable
decrease in background, because of the reduction in the annihilation
cross-section.  Thus, $\gamma\gamma$ physics at LEP2 is primarily an
extension of work done at LEP1, but we can hope to see an improvement in
our understanding as a result of larger data samples with less background
contamination, and in some cases useful hardware upgrades to the
experiments.

As can be seen from Professor Miller's review\cite{djm}, our present
understanding of the hadronic final state in $\gamma\gamma$ interactions
is not really satisfactory.  The underlying physics is a complex interplay
between soft (``vector dominance'') and hard (``direct'', ``QPM'',
``pointlike'')
QCD, with the added complication of variable photon virtuality.  Although
in recent years a number of Monte Carlo generators for $\gamma\gamma
\rightarrow \mbox{hadrons}$ have been produced\cite{HERWIG,PYTHIA,PHOJET},
none is claimed by its authors to be applicable to the whole experimental
range of $Q^2$, and none seems to provide a satisfactory description of
the experimental data even in its claimed range of validity.  The
situation is doubly unfortunate because of the need to ``unfold''
experimental $\gamma\gamma$ data to correct for the large detector
acceptance effects: if the event topology is not well reproduced by Monte
Carlo, the acceptance effects are unlikely to be well modelled, and so the
unfolding procedure becomes unreliable.  This results in large systematic
errors for one of the most important experimental measurements in
$\gamma\gamma$, the photon structure function $F_2(x,Q^2)$.

In view of the importance of this problem, both to $\gamma\gamma$ physics
and indeed to other LEP2 analyses where $\gamma\gamma$ interactions are a
significant background, the working group concentrated most of its efforts
on this question of modelling the hadronic final state.  Our aim was to
understand where the disagreements between simulations and data arise
from, and if possible to develop a prescription for ameliorating them.
The results of this study are presented in section 2 of this paper.
Section 3 considers in more detail the question of unfolding the photon
structure function, with particular reference to the possibility of
reducing systematic errors by unfolding in more than one variable.  This
approach relies on the fact that disagreement between Monte Carlo and data
{\it in the variable in which you unfold\/} is not important (if it were,
unfolding would be impossible, since the Monte Carlo would have to
incorporate the correct distribution of the unfolded variable---which is
what the unfolding is intended to discover).  Therefore, if one can find a
variable which characterises the difference between data and Monte Carlo
and unfold in that variable as well as in $x$, the systematic errors
caused by the discrepancy should be reduced.

It will be seen that both these sections are oriented towards the
inclusive production of hadrons in $\gamma\gamma$ interactions, and this
has indeed been the main focus of work at LEP1.  However, there is now
increased interest in exclusive final states, more specifically the
formation of meson resonances, because of the information that can be
gained on the possible glueball content of the meson so produced.  Lattice
gauge theory results strongly suggest that low-lying glueball states will
be close in mass to conventional mesons of the same $J^{PC}$, implying that
the observable states will be mixed.  A meson with a high gluonium content
should be suppressed in $\gamma\gamma$ production compared to gluon-rich
channels such as upsilon decay and central production in hadron
collisions.  Admittedly LEP2 is not an ideal environment for such studies,
because they impose stringent requirements on experimental triggers, but
the theoretical interest is such as to warrant a feasibility study.  This
is considered in section 4 of this paper.

\section{Modelling the hadronic final state}

\subsection{Introduction}

\begin{figure}[htb]
 \begin{minipage}[t]{0.47\linewidth}
 \epsfig{file=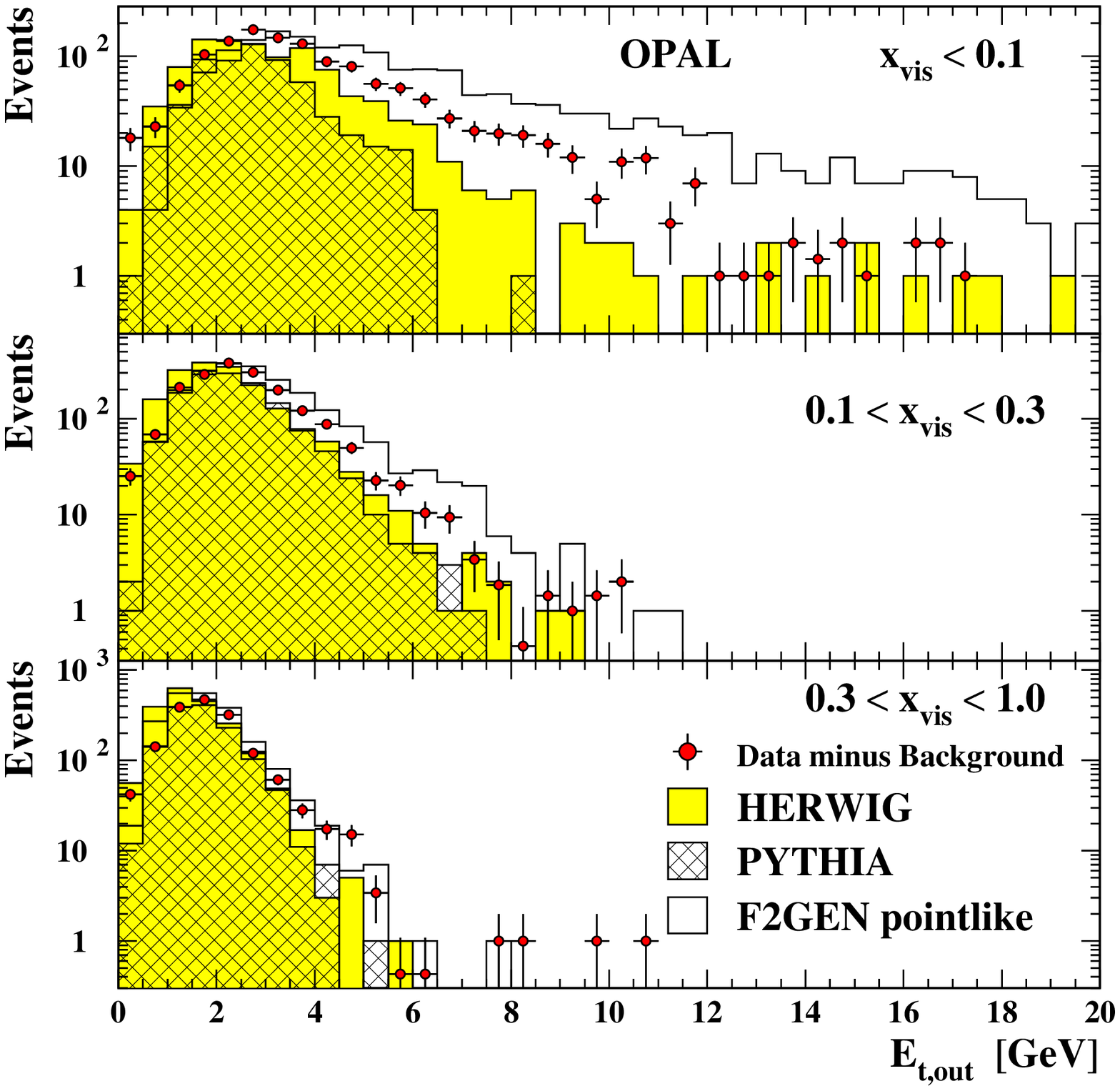,height=7.9cm,width=7.cm}
  \vspace{-0.1cm}
  \caption{\label{fig:ox_proc_01}
     Transverse energy out of the tag plane, which is defined by the tag
     direction and beam axis.}
 \end{minipage}\hfill
 \begin{minipage}[t]{0.47\linewidth}
  \epsfig{file=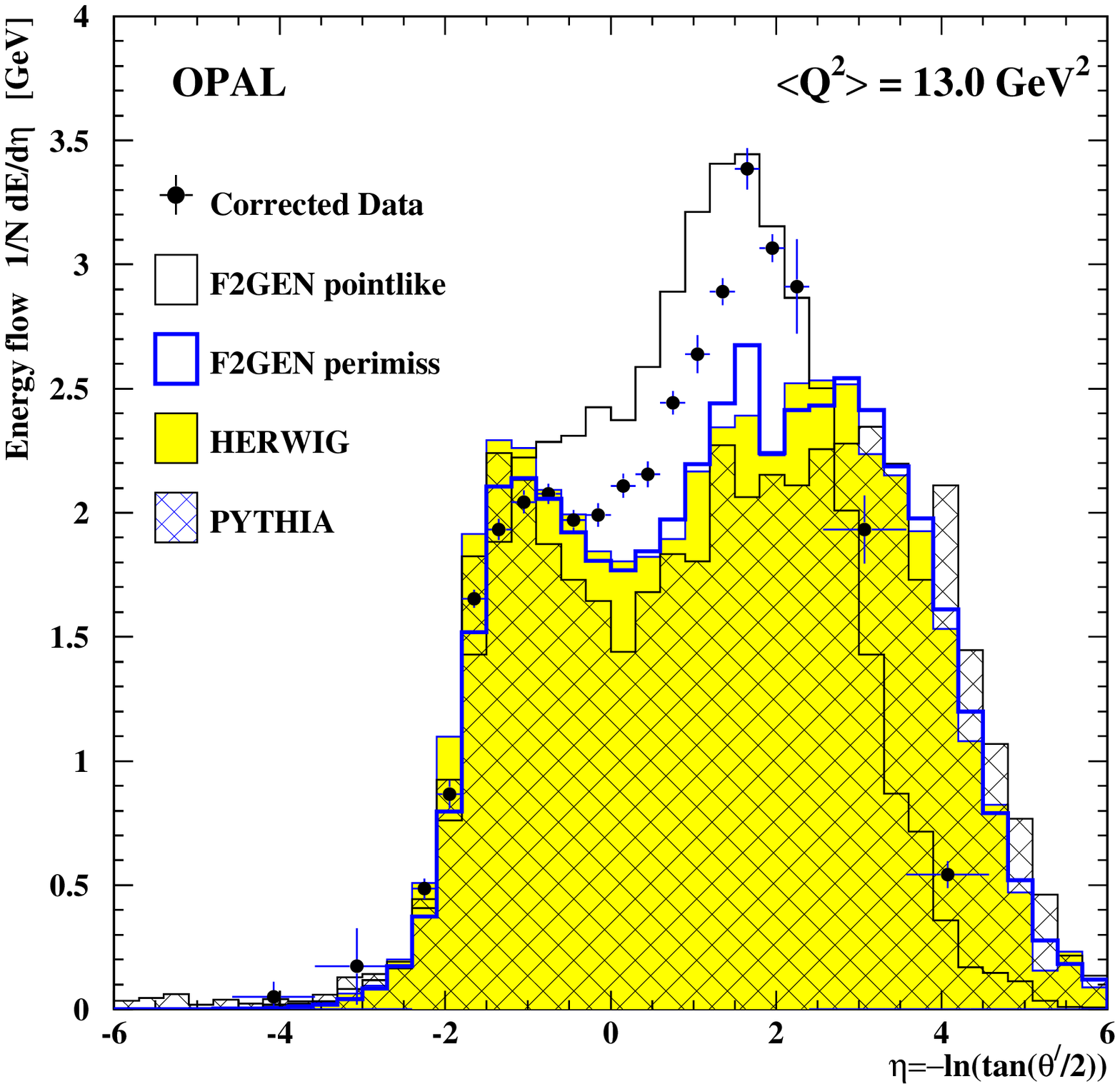,height=7.9cm,width=7.cm}
  \vspace{-0.1cm}
  \caption{\label{fig:ox_proc_02}
     Hadronic energy flow per event as a function of pseudorapidity,
     corrected for detector inefficiencies.}
 \end{minipage}
\end{figure}

 The measurement of $F_2^\gamma$ in deep inelastic $e\gamma$ scattering, where
 only one of the electrons is ``tagged'' in the detector and the other
 one escapes unseen,
 involves the determination of the  $\gamma^*\gamma$ invariant mass
 $W$ from the hadronic final state.
 Because of the non-uniform detection efficiency and incomplete
 angular coverage the correlation between $W_{\mathrm vis}$ and $W$ critically
 depends on the modelling of the hadronic final state.
 It has been shown~\cite{OPALPR185}
 that there exist serious discrepancies in the description
 of this hadronic final state.
 Fig.~\ref{fig:ox_proc_01} shows the transverse energy out of the plane,
 defined by the tag and the beam.
 For $x_{\rm vis}>0.1$ all of the generators are adequate,
 but for $x_{\rm vis}<0.1$ they are
 mutually inconsistent, and in disagreement with the data.
 At high \etout\ the data show a clear excess over \hwg~\cite{HERWIG}
 and \pyth~\cite{PYTHIA},
 while the pointlike \ftg~\cite{BUI-9401} sample exceeds the data.
 Similar discrepancies are
 observed in the hadronic energy flow per event~\cite{OPALPR185}, shown
 in Fig.~\ref{fig:ox_proc_02},
 where both \hwg\ and \pyth\ overestimate the energy in the forward
 region ($|\eta|>2.5$) and underestimate the energy in
 the central region of the
 detector.  At $|\eta|>2$ the data are closer to the pointlike distributions
 from \ftg\ than to the QCD models or the ``perimiss'' distributions
 of \ftg, a mixture of peripheral and pointlike
 events~\cite{BUI-9401}.
 But there is only one difference between the two \ftg\ samples;
 the angular distribution of the outgoing quarks in the $\gamma^\star\gamma$
 centre of mass system.
 This indicates that in tuning these models particular attention will need to
 be given to the parton distributions in the $\gamma^\star\gamma$
 system.

\subsection{Generating hadronic final states}

All the main Monte Carlo event generators for two photon physics were
described in detail in the LEP2 yellow book report\cite{yellow}.  Little
has changed since then, so we here only recap the salient features.  As
already mentioned, the angular distribution of the produced quarks is
crucial for a good description of the data, so we concentrate on how
this is generated in the different models.

\subsubsection{HERWIG}

In \hwg\ the backward evolution algorithm is used.  This means that the
scattering is first set up as $\mathrm{\gamma^*q \to q}$, with no
transverse momentum in the $\gamma\gamma^*$ frame.  Then the history of
the incoming quark is traced backwards in time to the target photon.  At
each step, the quark is `offered the chance' to have come from a quark
at higher $x$ via the emission of a gluon, $\mathrm{q \to qg}$ or
directly from a pointlike photon coupling, $\mathrm{\gamma \to
  q\bar{q}}$\footnote{In addition, it is offered the chance to have come
  from a gluon splitting, $\mathrm{g \to q\bar{q}}$, after which the
  gluon is evolved in the same way, coming from either a gluon at higher
  $x$, $\mathrm{g \to gg}$, or a quark, $\mathrm{q \to gq}$.}.  The
relative probabilities of these different steps are calculated from the
DGLAP evolution equation, i.e.\ effectively from the slopes in $x$ and
$Q^2$ of the chosen pdf set.  The evolution terminates either when a
pointlike photon coupling is generated (since the pdf of photons in
photons is a delta function at $x\!=\!1$), or when the evolution scale
has reached the infrared cutoff $\sim1$~GeV.  In the latter case, the
event is called hadronic, and in the former pointlike.  Thus this
separation is not made in advance but rather is generated dynamically,
and depends in a non-trivial way on the evolution of the chosen set of
parton distribution functions.  In the ideal world, this should
reproduce the assumptions made by the pdf fitters, but in practice there
are many non-idealities.  One finds\cite{Jeff} that \hwg\ calls many
more events hadronic than the pdf, particularly at small $x$.

This separation is of much more than passing interest, because it
determines the transverse momentum distribution of the remnant (or
`target') quark.  During this workshop, parton-level studies of several
event generators were made, which found that the emission of gluons
plays a relatively small r\^ole in determining the final state
properties, and that the most important factor is the distribution of
this target quark.  In pointlike photon events it receives a power-law
distribution according to the perturbative evolution equations, while in
hadronic events it receives a Gaussian distribution of adjustable width.
During the workshop we have discussed the effect of using other
distributions here, and show results below.

In addition to the backward evolution, \hwg\ is matched with the NLO
matrix elements, which ensure that the hardest emission is distributed
correctly.  This effectively includes the high-$p_t$ photon-gluon fusion
and pointlike photon processes, which can be particularly important at
small $x$.  They are not included in \pyth, which might account for the
fact that \hwg\ falls somewhat below the data, while \pyth\ falls
precipitously in Fig.~\ref{fig:ox_proc_01} for example.

\subsubsection{PYTHIA/ARIADNE}

\pyth\ has two options for its parton shower, either its own backward
evolution algorithm, or \ari's colour dipole cascade in which there
is no separation between initial-state and final-state emission.  In
either case it is decided in advance whether the event will be called
pointlike or hadronic, and the associated distributions generated
accordingly.

If \pyth's own algorithm is used, then in hadronic events the backward
evolution is similar to \hwg's described above, except that it is
never evolved back to a pointlike photon, i.e.\ a hadronic photon is
treated exactly like a hadron.  The remnant is given a Gaussian
transverse momentum distribution by default, although other shapes are
available, as discussed below.  In pointlike events, the evolution is
slightly different, deciding the transverse momentum of the remnant in
advance and using modified evolution equations that take that momentum
into account.  In common with \hwg, this transverse momentum
distribution is purely determined by perturbation theory, and there is
no freedom to adjust it in the default model.  During the workshop, we
have tried modifying this distribution by convoluting it with a narrow
(i.e.\ non-perturbative) Gaussian, and show results below.

If \ari's algorithm is used, the distribution of remnant momentum is
the same as in \pyth.  The evolution of the final state is modelled as
being from a colour dipole between the $\mathrm{q\bar{q}}$ pair, except
that the remnant quark is considered to be an extended object of size
$\sim1/p_t$.  This means that emission in the remnant direction with
$k_\perp \ltap p_t$ is suppressed.  Although \ari\ produces
significantly more gluon radiation than \pyth, particularly at small
$x$, the fact that we have found the distribution of remnant momentum to
be more important than of gluon momenta means that its predictions are
not significantly different from \pyth's.

\subsubsection{PHOJET}

\pho\ is an event generator aimed at a unified treatment of
$\mathrm{pp}$, $\gamma\mathrm{p}$ and $\gamma\gamma$ collisions.  It
provides a very complete picture of soft effects in these collisions
coupled with a somewhat less precise treatment of the perturbative
evolution.  It is aimed predominantly at the simulation of real photon
collisions and although it generates the electron vertex keeping track
of the target photon virtuality, it is only intended to be reliable for
relatively low virtualities.  Nevertheless, the OPAL collaboration have
found that if one ignores the warnings of the authors, and runs it at
higher virtualities anyway, one gets a fair description of
data\cite{Stefan}.  During this workshop, members of ALEPH have
tried the same comparison and found that good agreement could only be
attained by adding a vector meson form factor, and by reweighting the
$x$ and $Q^2$ distribution by hand as described below.  Having done
that, we find a good description of data.  One of the main features of
this model relative to \hwg\ and \pyth\ is the fact that it includes
contributions where both photons are resolved, even for $Q^2$ values
that would traditionally be described as deep inelastic scattering.  It
is possible that this is the main reason why the description of data is
improved.

\subsection{Studies carried out at the workshop}

\begin{figure}[htb]
 \begin{center}
 \vspace{-0.6cm}
 \epsfig{file=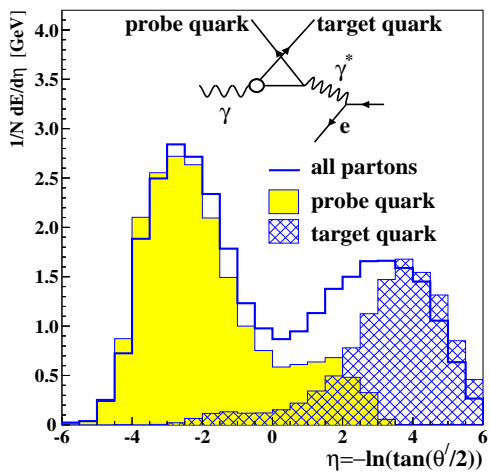,height=6.9cm,width=7cm}
 \vspace{-0.2cm}
 \caption{Energy flows of probe and target quarks.}
 \label{fig:ox_proc_03}
 \vspace{-0.1cm}
 \end{center}
 \end{figure}

 To study the contributions of the various partons, the \pyth/\ari~\cite{Leif}
 energy flow in the lab frame as a function of pseudorapidity
 is plotted  in Figure~\ref{fig:ox_proc_03}
 for the quark that couples to the off-shell
 probe photon $\gamma^\star$, denoted the probe quark, and for the
 quark that couples to the quasi-real target photon $\gamma$, denoted
 the target quark. The total energy flow
 of all partons after gluon radiation is also shown. The direction
 of the tagged electron is always at negative $\eta$.
 It is apparent that the hump at negative $\eta$ stems mostly from the
 probe quark which is scattered in the hemisphere of the tag, while the
 hump at positive $\eta$ originates mostly from the target quark
 in the opposite hemisphere of the struck photon.
 From comparisons of the hadronic energy flow of the data with the
 various models, it became apparent that the energy flow of the probe
 quark needs to be shifted to lower $\eta$, corresponding to an
 increased transverse energy. This can be achieved in
 several ways:

 Anomalous events carry more transverse momentum than hadronic events.
 Increasing the fraction of
 anomalous to hadronic or VMD type events would have the desired effect,
 but the pdf sets used (in this case
 SaS1D~\cite{SCH-9501} in \pyth\ and GRV~\cite{GLU-9202}
 in \hwg) do not readily allow changing this ratio.

 Another way to increase the transverse energy is to allow for more
 gluon radiation. This can be achieved by augmenting the inverse
 transverse size of the remnant, $\mu$, in the \ari\
 colour dipole model. In standard \ari, $\mu$ is set proportional to
 the intrinsic $k_T$ of the
 struck quark on an event-by-event basis.  For VMD events,
 $k_T$ is Gaussian with a width of 0.5~GeV.  For anomalous events $k_T$
 follows a power law.  But even a generous increase of the $\mu$
 parameter ($\mu=10$)
 has a relatively small effect on the partonic energy flow.

\begin{figure}[htb]
 \vspace{0.5cm}
 \begin{center}
  {\epsfig{file=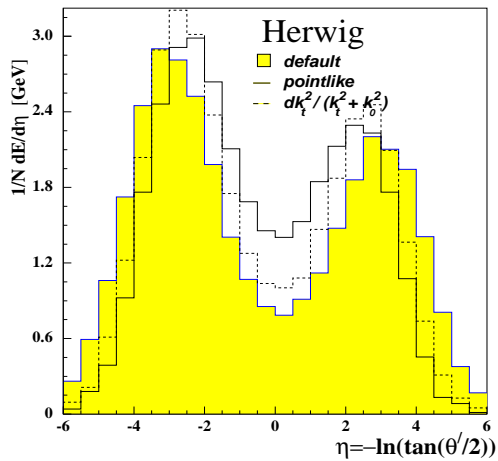,height=6.9cm,width=0.55\linewidth}}
 \vspace{-2.1cm}
  \caption{\label{fig:ox_proc_04} \hwg\ energy flow of events generated
           with different $k_T$ distributions.}
 \hspace{-0.9cm}
 \vspace{-0.4cm}
 \end{center}
\end{figure}

 Increasing the intrinsic transverse momentum $k_T$ of the partons in the
 struck photon
 is another way of directly influencing the angular distribution of the
 hadronic final state.   Figure~\ref{fig:ox_proc_04} shows the energy flow
 for \hwg\ events with default settings, with a fully pointlike
 distribution and with an enhanced $k_T$ distribution.  Note that the
 run with a fully pointlike distribution looks pretty similar to the
 \ftg\ all pointlike results, a non-trivial cross check that, in
 pointlike events, \hwg\ seems to be behaving as expected, it is just
 that it is not producing enough pointlike events.  As a fix, the
 enhanced $k_T$ distribution appears to be the most promising.

 \vspace{-0.5cm}

\begin{figure}[hbt]
 \begin{center}
 \vspace{-0.2cm}
 \begin{minipage}[t]{0.35\linewidth}
    \epsfig{file=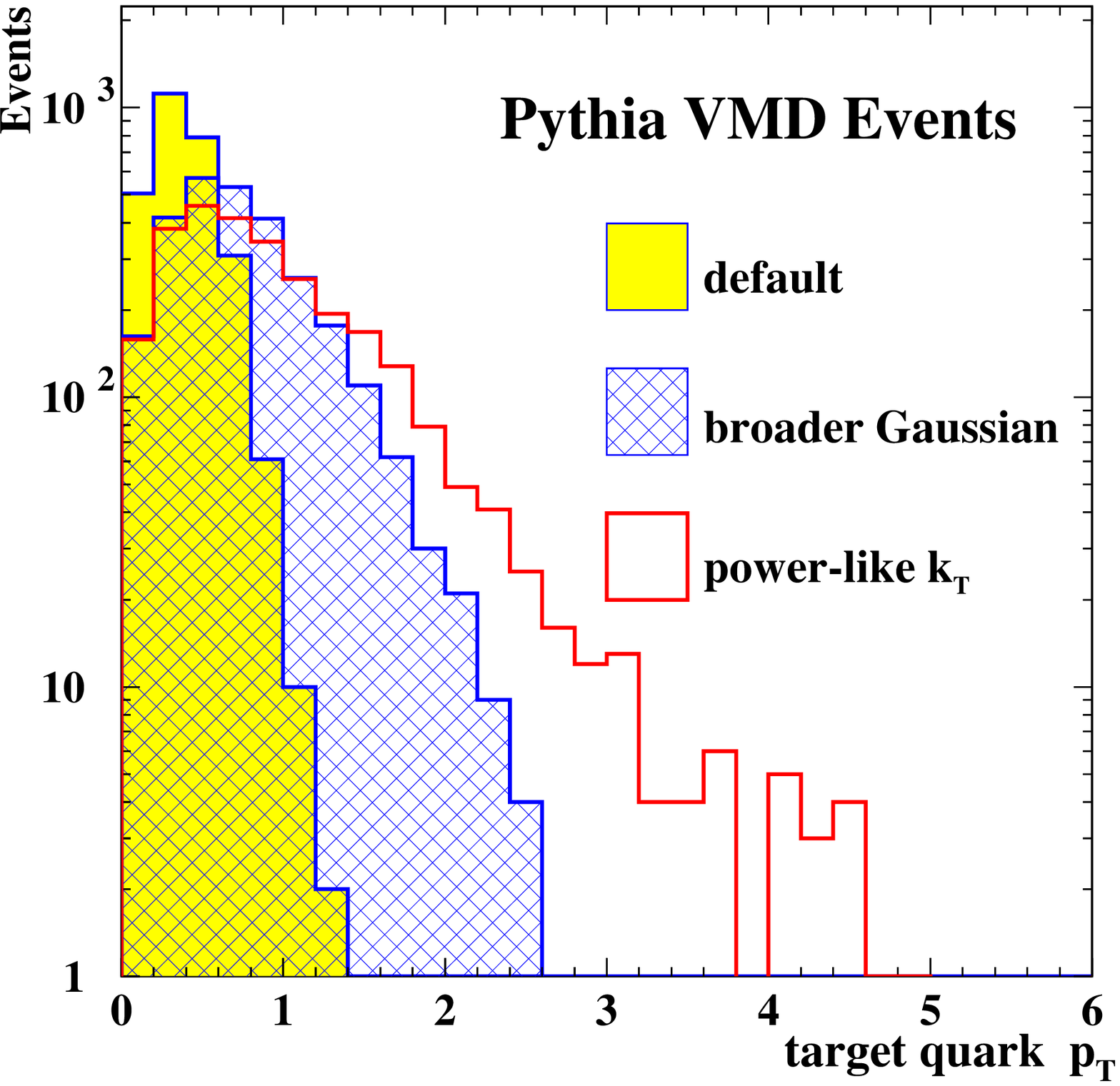,height=5.cm,width=5.5cm}
 \end{minipage}
   \hfil
 \begin{minipage}[t]{0.35\linewidth}
    \epsfig{file=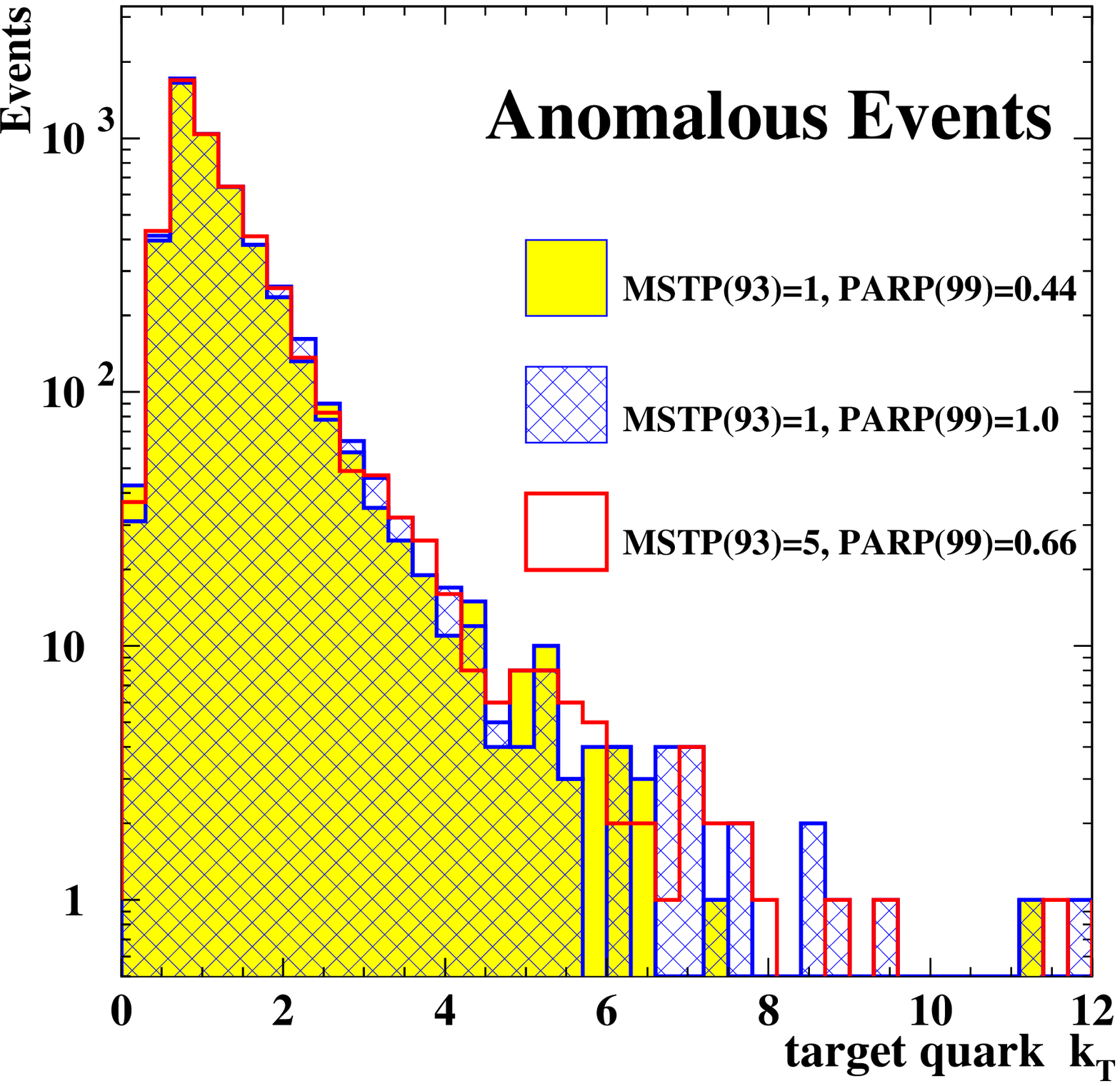,height=5.cm,width=5.5cm}
 \end{minipage}
 \vspace{-0.2cm}
 \caption{\label{fig:ox_proc_05} Transverse momentum of the target quark
                              in \pyth/\ari~for VMD and anomalous events.}
 \end{center}
 \end{figure}

 In \pyth\ the pdf determines whether an event is generated as a VMD
 or an anomalous event.
 The intrinsic $k_T$ of the quasi-real
 photon can be controlled with parameters~\cite{PYTHIA}.
 Just increasing the width of the Gaussian distribution does not produce
 events that populate the region of high \etout\ at low $x$ observed
 in the data (Fig.~\ref{fig:ox_proc_01}).  A similar deficiency had been
 observed in the resolved photoproduction data at ZEUS~\cite{ZEUS-B354},
 which led to the introduction of a
 power-like $k_T$-distribution of the form \dkt ,
 improving the distributions of the photon remnant. The parameter $k_0$
 is a constant, set to 0.66 GeV in~\cite{ZEUS-B354}.
 The \pyth\ parameters only allow adjusting the $k_T$ for VMD type events,
 figure~\ref{fig:ox_proc_05}, but not for anomalous evens.
 To change the intrinsic $k_T$ of anomalous events a
 Gaussian smearing is added in quadrature~\cite{Torbjorn}.

\subsection{Comparisons of models with data}

\begin{figure}[htb]
\vspace{-0.5cm}
 \begin{minipage}[t]{0.47\linewidth}
  \epsfig{file=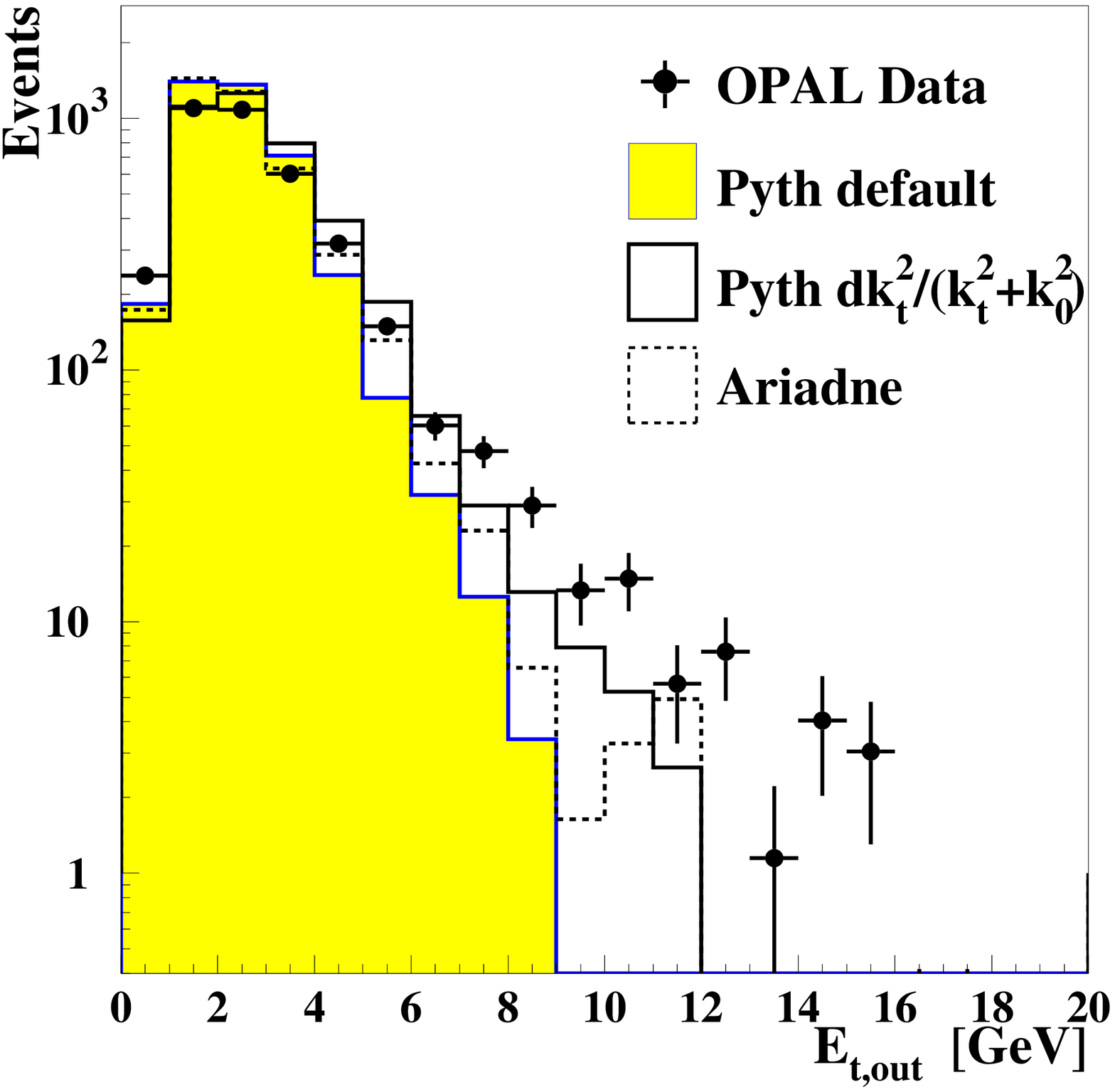,height=6.1cm,width=6.9cm}
  \vspace{-0.1cm}
  \caption{\label{fig:ox_proc_06} \pyth :
     Transverse energy out of the tag plane.}
 \end{minipage}\hfill
 \begin{minipage}[t]{0.47\linewidth}
  \epsfig{file=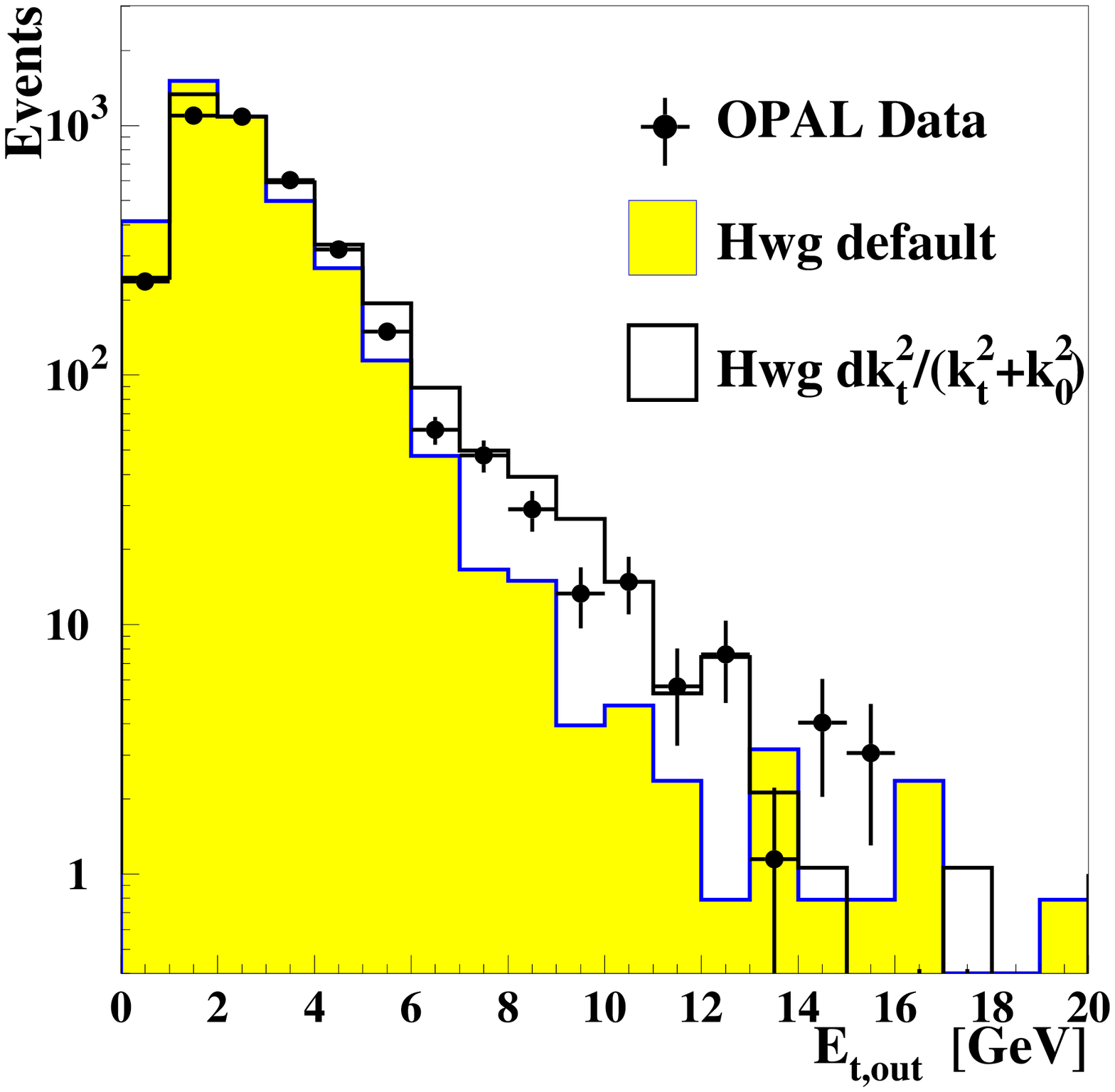,height=6.1cm,width=6.9cm}
  \vspace{-0.1cm}
  \caption{\label{fig:ox_proc_07} \hwg :
     Transverse energy out of the tag plane.}
 \end{minipage}
\end{figure}

\begin{figure}[htb]
 \begin{minipage}[t]{0.47\linewidth}
  \epsfig{file=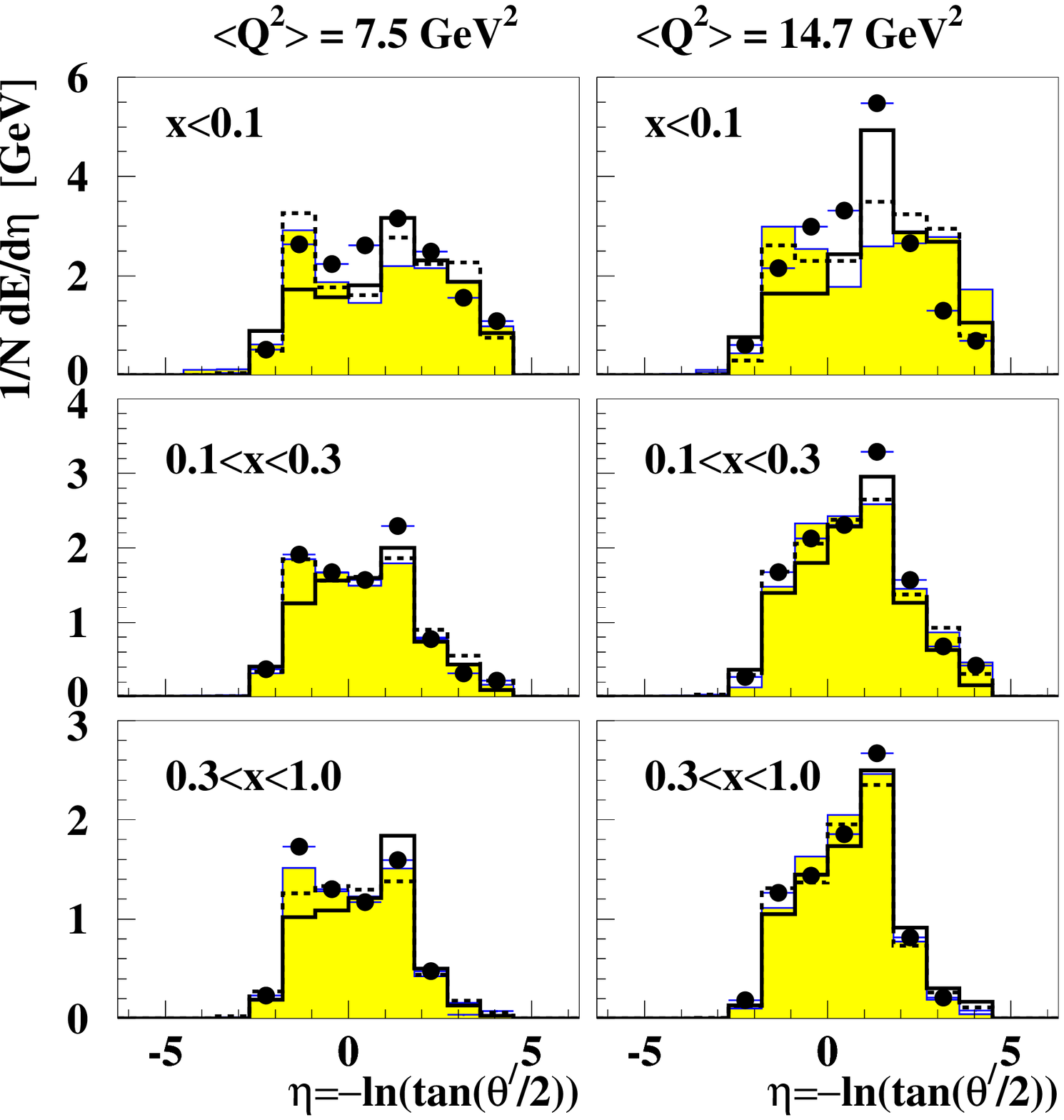,height=6.5cm,width=6.5cm}
  \vspace{-0.1cm}
  \caption{\label{fig:ox_proc_08} $\!$\pyth :
     Hadronic energy flow as function of $x$ and $Q^2$.
     The symbols are the same as in Fig. 6}
 \end{minipage} \hfill
 \begin{minipage}[t]{0.47\linewidth}
  \epsfig{file=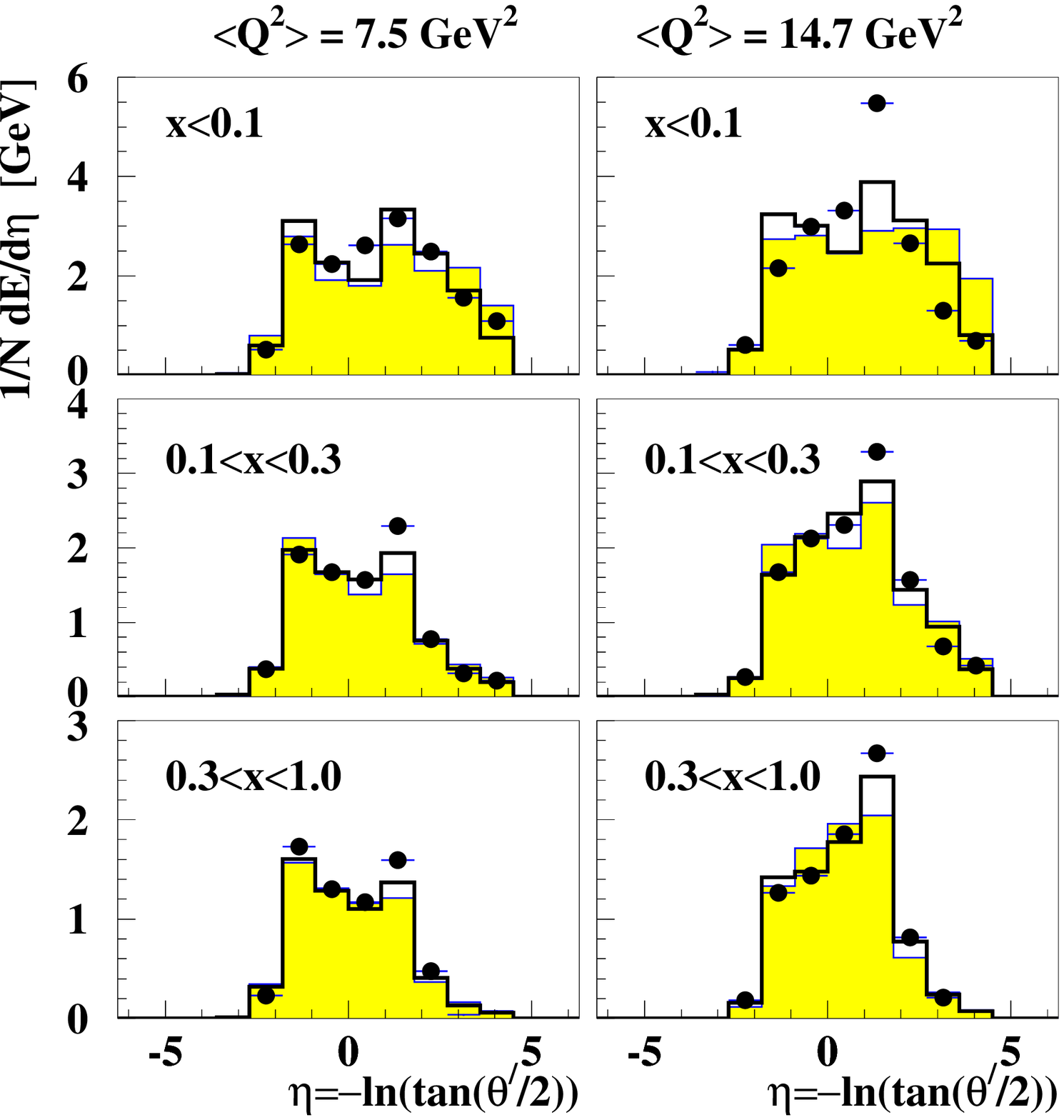,height=6.5cm,width=6.5cm}
  \vspace{-0.1cm}
  \caption{\label{fig:ox_proc_09} $\!$\hwg :
     Hadronic energy flow as function of $x$ and $Q^2$.
     The symbols are the same as in Fig. 7}
 \end{minipage}
\end{figure}

 Figure~\ref{fig:ox_proc_06} show the \etout\ and
 figure~\ref{fig:ox_proc_08}
 the hadronic energy flows on detector level of \pyth\ with default
 parameter settings and with the \dkt\ distribution for VMD
 plus a Gaussian smearing
 of the anomalous events, compared to the OPAL data taken
 in 1993--1995 at $\sqrt{s_{ee}}=91$ GeV~\cite{jal}.
 In addition the \ari\ distributions
 with enhanced gluon radiation are shown.   While the \etout\ spectrum
 has been improved, it still falls short of the data in the tail of
 the distribution.  The hadronic energy flow generated by the enhanced
 \pyth\ recreates the peak on the remnant side (positive $\eta$) seen
 in the data at low $x_{\rm vis}$, at the expense of a somewhat worse fit
 on the tag side.


 \hwg\ separates events dynamically into hadronic and anomalous type.
 A similar \dkt\ distribution of the intrinsic transverse momentum
 of hadronic photons can be added by hand.  The results of this are
 shown in figures~\ref{fig:ox_proc_07} and \ref{fig:ox_proc_09}.  Both the
 \etout\ and the energy flows are greatly improved
 with the inclusion of the power-like $k_T$ distribution, with the
 exception of the peak in the energy flow at low $x_{\rm vis}$ -- high
 $Q^2$, which still falls short of the data.

\begin{figure}[htb]
 \begin{minipage}[t]{0.47\linewidth}
  \epsfig{file=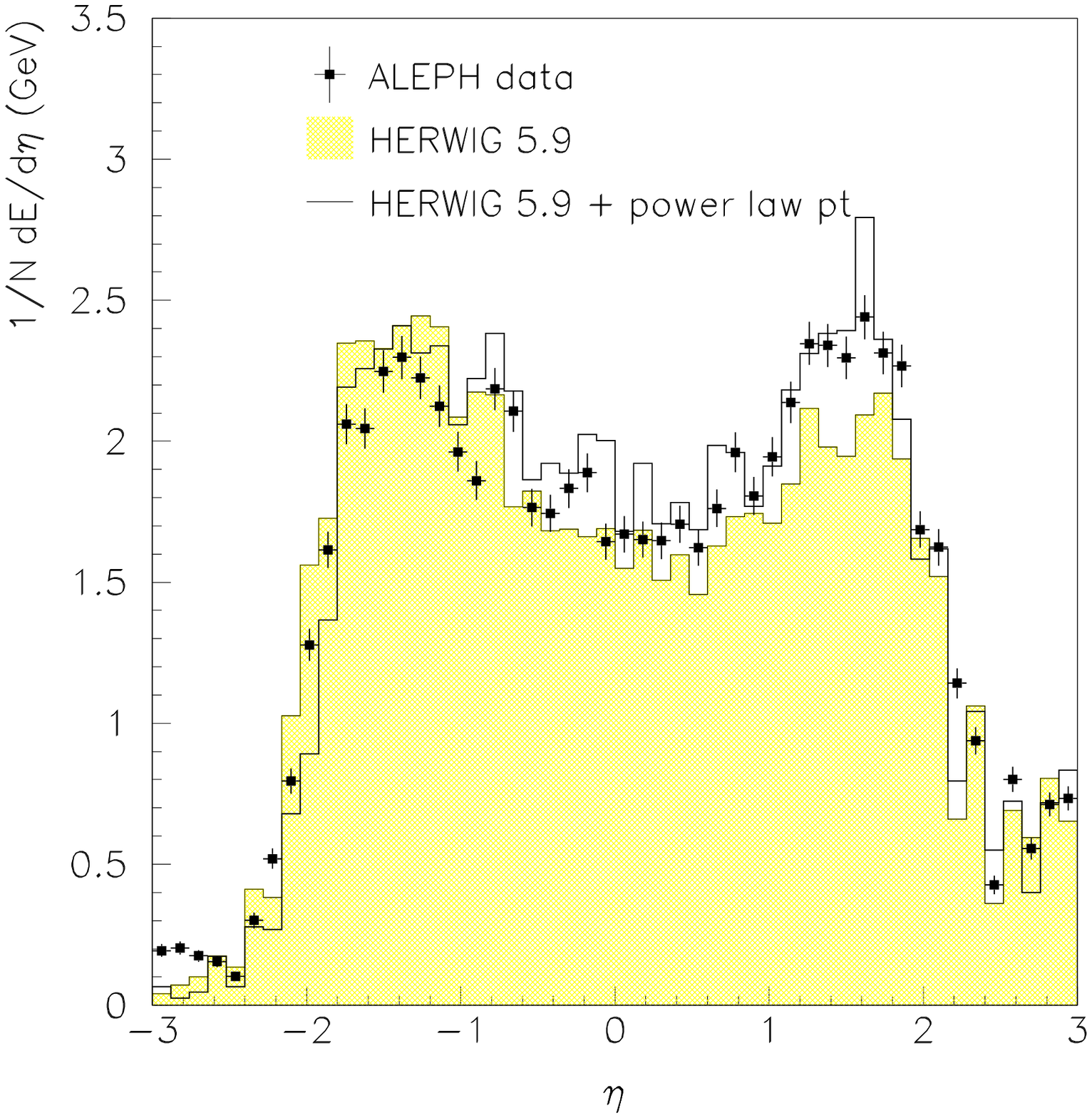,width=6.5cm}
  \vspace{-0.1cm}
  \caption{\label{aleph1}Hadronic energy flow: \hwg\ vs.\ ALEPH.}
 \end{minipage} \hfill
 \begin{minipage}[t]{0.47\linewidth}
  \epsfig{file=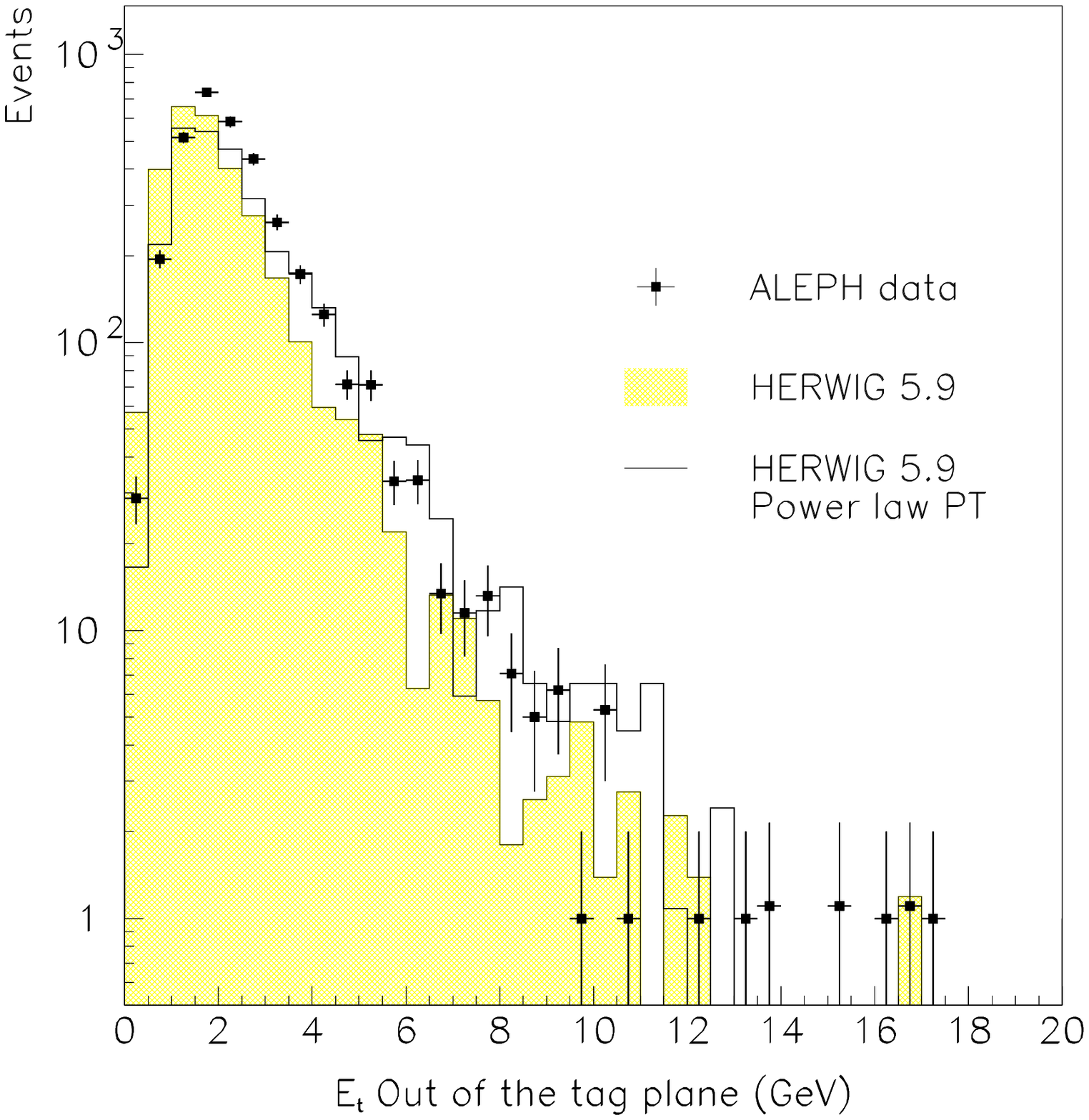,width=6.5cm}
  \vspace{-0.1cm}
  \caption{\label{aleph2}Transverse energy out of the tag plane: \hwg\
  vs.\ ALEPH.}
 \end{minipage}
\end{figure}

A similar improvement is also seen in the description of ALEPH
data\cite{alj}, as seen in figures~\ref{aleph1} and~\ref{aleph2}.  In
these plots the average $Q^2$ is 14.2~$\gev^2$ and the full $x$ range is
integrated over.


 Overall, the power-like distribution of the intrinsic transverse momentum
 of the struck photon of the form \dkt\ greatly improves the hadronic
 final state distributions of both \pyth\ and \hwg . This improved
 description of the data should reduce the model-dependent systematic
 errors in the unfolded result of the photon structure function
 $F_{2}^{\gamma}$. More fine-tuning of these models is required.


A number of LEP experiments have observed that the \pho\ program provides the
best available description of real photon interactions as observed in untagged
events. Despite this being expressly forbidden by the author of the
program, it has
also been compared to tagged events by OPAL and found to give an acceptable
description of their data\cite{Stefan}. At the workshop this was attempted
for the first
time using ALEPH tagged data.

\begin{figure}[tb]
\centering
\leavevmode
\epsfig{file=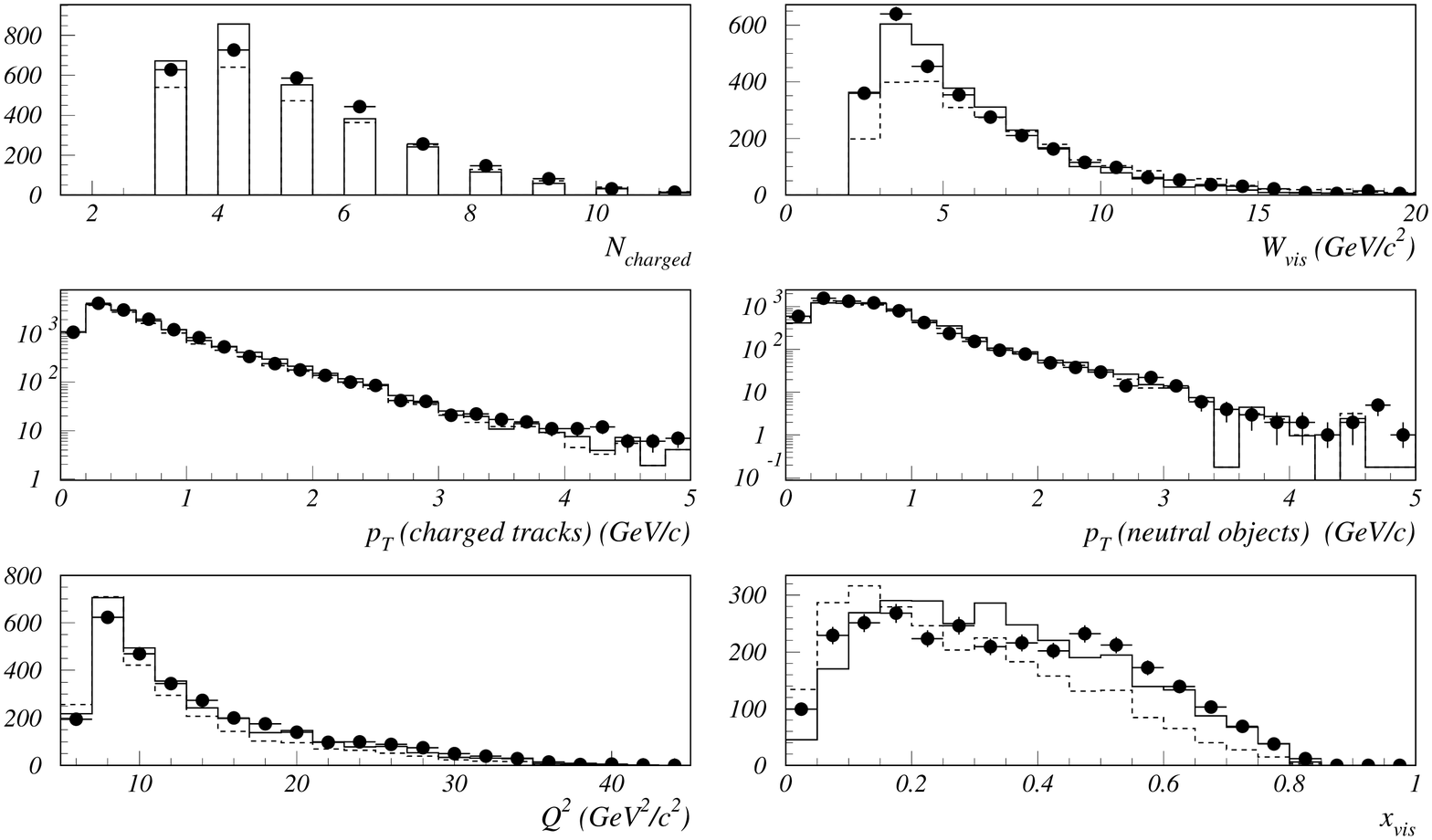,height=7cm,width=1.0\textwidth}
\caption{\label{fig_MC}Comparison of \pho\ to the data.
The data are shown by the points with error bars. The
histograms are the sum of the \pho\ and  $\gamma \gamma \rightarrow \tau^+\tau^-$
models. The \pho\ model has been normalised to the data. The solid line is the
result after reweighting the events to obtain the same $x_{true}$ distribution
as is found in \hwg\ with GRV input structure function. The dashed line is the
Monte Carlo data before this weighting has been applied.
 }
\end{figure}

\begin{figure}[tb]
\centering
\leavevmode
\epsfig{file=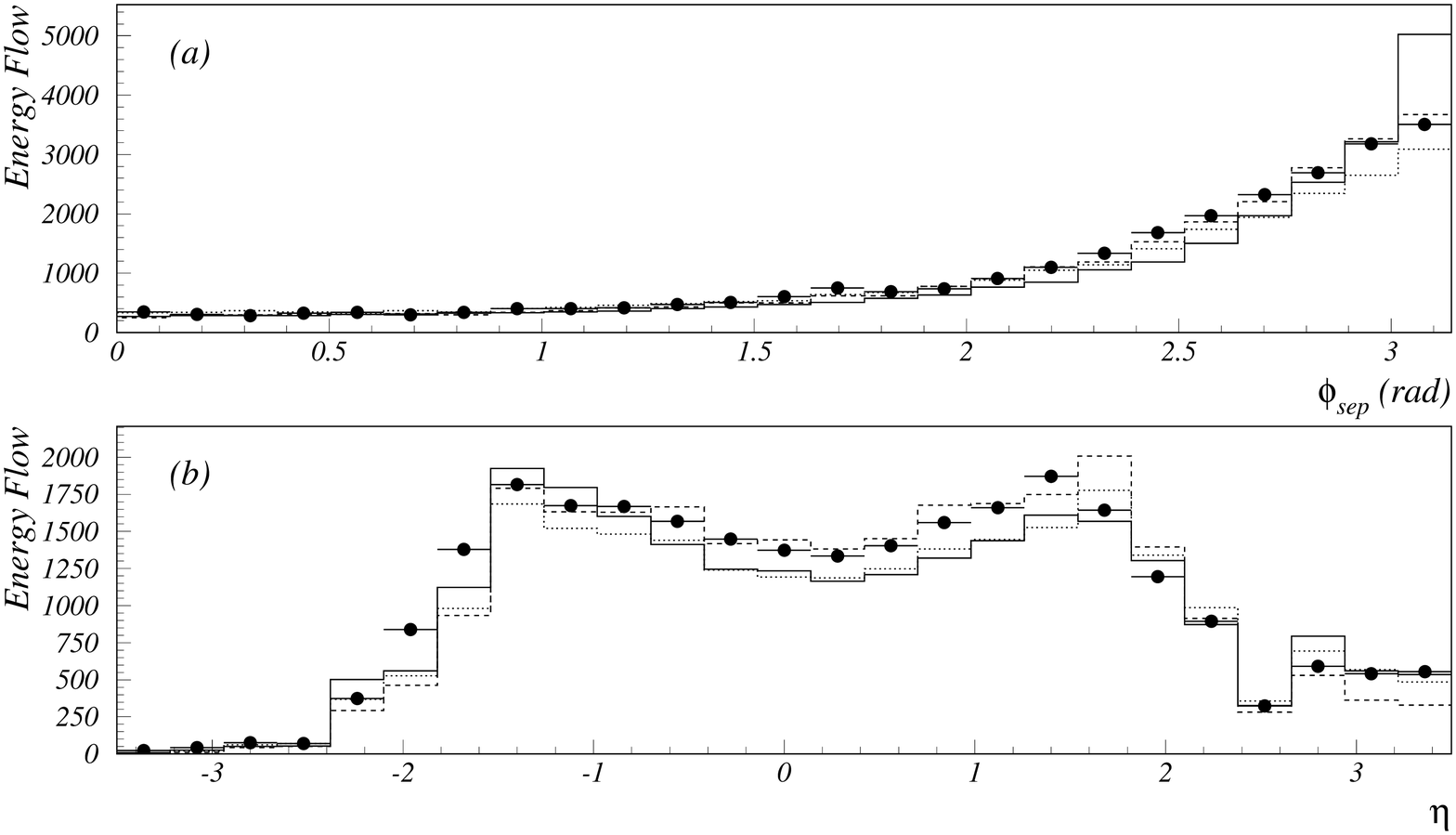,height=7cm,width=1.0\textwidth}
\caption{\label{fig:eta_phi} Energy flow as a function of (a) \phiSep, (b)
$\eta$.
The points are the data. Solid line:  \hwg\ with GRV;
dotted line: \pho; dashed line: reweighted
\pho.}
\end{figure}

The event selection is described in Ref.~\cite{Gordon}. The ALEPH version of
\pho\ was adapted to produce tagged events with a Generalized Vector
Dominance form factor. A comparison of the model to the
data is given in figures \ref{fig_MC} and \ref{fig:eta_phi}.  The dashed lines
in these distributions show the \pho\ distributions that resulted.
In some of the
distributions the agreement between data and Monte Carlo is better than that
achieved by existing models. However the $x_{vis}$ distribution shows the Monte Carlo
exceeding the data at low $x_{vis}$ which suggest that the $x_{true}$
distribution is also too strongly peaked at low values. This could also cause
the disagreement in the $W_{vis}$ and $Q^2$ distributions. In order to test
this the true $x$ distribution was compared for \pho\ and \hwg\ using the GRV
structure function as input. This version of \hwg\ gives a good description of
the ALEPH $x_{vis}$ distribution in this $Q^2$ region. The ratio of these two
$x_{true}$ distributions was obtained and parameterised with
a fourth degree polynomial. This function was then used to  weight the events
in the histogram. This is shown by the solid line in figure \ref{fig_MC}
and results in a better fit to the data in the variables considered here
than the models currently used in ALEPH.

As we have seen several times at this workshop, the models currently
used in tagged $\gamma^* \gamma$ studies do not describe
the  data  when considered in terms of the pseudorapidity
 $\eta$ and and the azimuthal separation   $\phi_{sep}$.
  Fig.~\ref{fig:eta_phi}  shows these distributions for
the ALEPH
data compared to the Monte Carlo events. It can be seen that the tagged \pho\
model does a better job of fitting the data than does \hwg, and this
improvement is even greater once the reweighting has been applied.

\subsection{Summary and conclusions}

It is becoming steadily clearer and clearer that the data show a much
more pointlike structure than the QCD-inspired Monte Carlo models.  In
order to provide reliable unfolding of the structure function $F_2$, it
is essential to have theoretically well-founded models that are capable
of giving good fits to data after variation in $F_2$.  We do not have
any models that fulfil both criteria at present.  They must be
well-founded, because we must know that the $x_{\mathrm{vis}}$
distribution extracted is somehow related to that predicted
theoretically.  They must be able to give a good fit to data, otherwise
we will not believe the unfolding procedure at all.  One way out of this
situation is to unfold in several additional variables simultaneously,
thereby unfolding away the model-dependence, as discussed in the next
section.  Another is clearly to improve the models until they do
fulfil the criteria.

While we have come no closer to finding a theoretically-sound model that
describes data well, we have made progress in finding models that are
able to give a good description of data at all.  Unfortunately this has
meant poorly-motivated modifications in each case: in \pyth\ we have
added Gaussian intrinsic transverse momentum to anomalous events; in
\hwg\ we have added power-like transverse momentum to hadronic events;
and we have used \pho\ in regions for which it should not be valid.

We should not however be entirely negative about these modifications.
Two photon physics is not out on its own, but is closely related to
other areas of particle physics, particularly those being studied at
HERA, proton DIS and photoproduction.  Similar problems are being found
there~-- too little transverse momentum in the proton direction in DIS
and in the photon remnant direction in photoproduction.  We can
therefore draw some phenomenological satisfaction from the fact that the
same solutions seem to work in both cases, like including a power-like
photon remnant transverse momentum distribution and a resolved component
in DIS.

\section{Unfolding the photon structure function}

The large integrated luminosity expected from the LEP II programme
will provide new opportunities to investigate $\gamma \gamma$ collisions.
Single-tag events, where one electron remains in the beam pipe and the other
is measured with an angle $\theta_{\rm tag}$ and energy $E_{\rm tag}$, can be
viewed as the deep inelastic scattering of an electron and an (almost) real
target photon.  Of particular interest is the joint distribution of the
negative four-momentum squared of the probing photon,
$Q^2 = 4 E_{\rm tag} E_{\rm beam} \sin^2 (\theta_{\rm tag}/2)$, and the
invariant mass of the hadronic system, $W$.  Equivalently, one usually
measures the distribution $Q^2$ and the variable $x = Q^2 / (Q^2 + W^2)$;
this can be directly related to the photon structure function
$F_2^{\gamma}(x,Q^2)$. (The formalism of single-tag $\gamma \gamma$
collisions is described in e.g.\ \cite{singtag}).  In this paper we
concentrate on the question of measuring the distribution of $x$
for single-tag $\gamma \gamma$ collisions within a given narrow
range of $Q^2$.

The dominant uncertainties in current measurements of the $x$ distribution
are related to corrections that must be introduced to account for
finite acceptance and resolution of the detector.  These effects stem mainly
from hadrons at low angles with respect to the beam line that escape
detection.  This results in a lower measured hadronic mass compared to
its true value, and hence in a corresponding distortion of the variable $x$.
This is in contrast to the situation with $Q^2$, which is entirely
determined by the tag electron's angle and energy.  It can therefore
be measured to within several percent, and corrections
for resolution effects do not pose a major problem.  Here we will
concentrate on unfolding the distribution of $x$.

A detailed description of unfolding problems can be found in \cite{glenstat}.
We can represent a sample of measured values of $x$ by means of a
histogram with $M$ bins.  The expectation values of the numbers of events
$\vec{\mu} = (\mu_1, \ldots, \mu_M)$ that would be obtained with
a perfect detector are thus the parameters which we want to estimate.
What we obtain from the experiment is a histogram
$\vec{n} = (n_1, \ldots, n_N)$, which is distorted with respect to
$\vec{\mu}$ both because of statistical fluctuations as well as
from the effects of acceptance and resolution.  The latter can cause
an event with a true value of $x$ in bin $i$ to be observed in some
different bin $j$.  (Note that in general the $M$ bins for the true
histogram $\vec{\mu}$ need not be the same as the $N$ bins of the
observed histogram $\vec{n}$.  In the following, however, we will use the
same binning for both.)

The number of entries $n_i$ observed in a given bin $i$ can be treated as
a Poisson variable with expectation value $\nu_i = E[n_i]$.
The vectors $\vec{\mu}$ and $\vec{\nu}$ are related by

\begin{equation}
\label{mu_to_nu}
\vec{\nu} = R \vec{\mu} \, + \, \vec{\beta} \:,
\end{equation}

\noindent where the matrix element $R_{ij}$ represents the probability
for an event to be observed in bin $i$ given that its true $x$ value
was in bin $j$, and the vector $\vec{\beta} = (\beta_1, \ldots, \beta_N)$
gives the expected background.  Here for simplicity we
will neglect the background, and thus equation (\ref{mu_to_nu}) becomes
$\vec{\nu} = R \vec{\mu}$.  Note that a true value in bin $j$ need not
lead to any measured value at all, i.e.\ the efficiencies

\begin{equation}
\label{efficiency}
\varepsilon_j = \sum_{i=1}^N R_{ij}
\end{equation}

\noindent are in general less than unity.

If we had the vector $\vec{\nu}$, then we could simply invert $R$ to obtain
$\vec{\mu} = R^{-1} \vec{\nu}$.  What we have instead, however, are the data
values $\vec{n}$, which are subject to random fluctuations.  The estimators
$\vec{\hat{\mu}} = R^{-1} \vec{n}$ are unbiased, but have extremely
large variances.  (In the following, estimators will be denoted
by hats.)  These are in fact the estimators that one obtains by maximizing
the log-likelihood function based on Poisson-distributed data,

\begin{equation}
\label{log_L}
\log L(\vec{\mu}) = \sum_{i=1}^N \log \left(
\frac{\nu_i^{n_i}}{n_i!} \, e^{-\nu_i} \right) \:,
\end{equation}

\noindent which becomes

\begin{equation}
\label{log_L2}
\log L(\vec{\mu}) = \sum_{i=1}^N (n_i \log \nu_i - \nu_i)
\end{equation}

\noindent after dropping terms not depending on the parameters.  (Note
that this is regarded as a function of $\vec{\mu}$, since one has
$\vec{\nu} = R \vec{\mu}$.)

The idea behind unfolding is to construct estimators $\vec{\hat{\mu}}$ with
much smaller variances than the maximum likelihood estimators, at the
cost of introducing a small bias.  In regularized unfolding,
this is done by choosing the smoothest solution (according to some criterion)
out of those for which the log-likelihood is within some $\Delta \log L$ of
its maximum value.  This is equivalent to maximizing a linear combination of
$\log L(\vec{\mu})$ and a regularization function $S(\vec{\mu})$,

\begin{equation}
\label{phi}
\varphi(\vec{\mu}, \lambda) =
\alpha \log L(\vec{\mu}) \, + \, S(\vec{\mu}) \, + \,
\lambda \left[ n_{\rm tot} \, - \, \sum_{i=1}^{N} \nu_{i}\right] \:,
\end{equation}

\noindent with respect to the parameters $\vec{\mu}$ and the Lagrange
multiplier $\lambda$.  The function $S(\vec{\mu})$ must be defined to reflect
the smoothness of the solution.  The regularization parameter $\alpha$
determines the trade-off between likelihood and smoothness, and can be chosen
to correspond to a given value of
$\Delta \log L = \log L_{\rm max} - \log L(\vec{\mu})$.  The final term
in $\varphi$ restricts the solution to satisfy
$\sum_i \nu_i = \sum_{i,j} R_{ij} \mu_j = n_{\rm tot}$, where $n_{\rm tot}$
is the total number of events observed.

A commonly used regularization function is based on the mean squared
second derivative of the unfolded distribution (Tikhonov regularization).
Computer programs based on this technique have been used in previous
structure function measurements \cite{Blobel,SVD}.
The regularization function can be implemented by approximating the
second derivative of the distribution using finite
differences.  This leads to a function of the form

\begin{equation}
\label{Tikhonov}
S(\vec{\mu}) = - \sum_{i,j=1}^M G_{ij} \, \mu_i \, \mu_j \:,
\end{equation}

\noindent where coefficients $G_{ij}$ must be determined so as to
take into account differences in the bin widths; cf.\ \cite{glenstat}.
Another possible regularization function is

\begin{equation}
\label{entropy}
S(\vec{\mu}) = - \sum_{j=1}^M \mu_j \log \frac{\mu_j}{\mu_{\rm tot}}
= - \mu_{\rm tot} \sum_{j=1}^M p_j \log p_j = \mu_{\rm tot} H \:,
\end{equation}

\noindent where $\mu_{\rm tot} = \sum_{i=1}^M \mu_i$,
$p_j = \mu_j / \mu_{\rm tot}$, and $H$ is the Shannon entropy.
The entropy-based regularization function makes no
reference to the relative locations of any of the bins, which has the
advantage that the bins at the edges of the distribution are treated on the
same footing as those in the middle.  In addition, use of (\ref{entropy})
(referred to in the following as MaxEnt) can be directly applied to
multidimensional distributions.

One can show that the estimators $\vec{\hat{\mu}}$ obtained from any
regularized unfolding technique are biased.  An additional systematic error
is related to the model dependence of the response matrix $R$.  This matrix
must in general be determined by a Monte Carlo calculation where events are
generated and processed by a detector simulation program.  By construction,
the conditional probability for an event to be observed at $x'$ given that it
was generated at $x$ is independent of the distribution of $x$.
The corresponding statement for finite bins in $x$ is approximately true
as long as the bins are not too large.

The true value of $x$ is not, however, the only variable that has an
influence on the probability to measure a given value $x'$.  For example, if
the hadrons are mostly at low angles with respect to the beam line, then the
resolution for $x$ will be poor, since on average more particles
will be lost.  Since different models have in general different distributions
for all of the variables that characterize the final state hadrons, each
will lead to somewhat different response matrices $R$, and hence to different
results for the unfolded $x$ distribution.

A method to reduce this model dependence is to measure for each event not
only $x$, but in addition some other variable $y$ (to be defined) that also
characterizes the final state.  The response matrix gives the probability for
an event with true values of $x$ and $y$ in bins $i$ and $j$ to be observed in
bins $k$ and $l$.  This matrix is now by construction independent of the
model's joint distribution of $x$ and $y$.  In principle this idea could be
extended to an arbitrary number of additional variables, and eventually
one would eliminate all model dependence from the response matrix.  Since
the individual cells of the multidimensional space must be populated with
enough Monte Carlo events to determine the response matrix, however, the total
number of bins, and hence the number of variables, is necessarily limited.
Here we will only consider the case of two measured variables.

In order to reduce the model dependence of $R$, the second variable
$y$ should be related to the detector's ability to measure $x$.  Here,
we consider $y = |\cos \theta_{\rm had}|$, where
$\theta_{\rm had}$ is the polar angle of the total hadronic system in
the laboratory frame.  In fact, large discrepancies between observed and
predicted angular distributions of hadrons have been observed in
single-tag events \cite{OPAL}.  Thus some reduction in the
uncertainty from model dependence can be expected by obtaining
this information directly from the data.  Other possibilities
are the orientation of the hadronic system with respect to the $\gamma \gamma$
axis, or a quantity related to hard gluon radiation, such as the thrust
or jet structure of the hadrons measured in the $\gamma \gamma$ rest frame.
After unfolding the distribution of $x$ and $|\cos \theta_{\rm had}|$,
we will integrate over $|\cos \theta_{\rm had}|$ to obtain the distribution
of $x$ alone.

In order to investigate these ideas quantitatively, single-tag
$\gamma \gamma$ events were generated with the \hwg\ Monte Carlo
generator \cite{HERWIG}.  This allows the user to choose from
a variety of parametrizations for the photon structure function
from the package PDFLIB \cite{PDFLIB}.

The response matrix was determined by means of a highly simplified
detector simulation program, roughly corresponding to a typical
LEP detector.   An electromagnetic calorimeter was assumed to be sensitive to
photons and electrons with energies $E > 200$ MeV, to be hermetic down
to 30 mrad from the beam line, and was taken to have an energy resolution of
$\sigma_E = 0.18 \sqrt{E}$.  A hadron calorimeter in the angular range
$|\cos \theta | < 0.99$ was assumed to be fully efficient for energies
above 1.0 GeV, and to have an energy resolution of $\sigma_E = 0.80 \sqrt{E}$.
Charged particles with $|\cos \theta | < 0.94$ and transverse momentum
$p_{\perp} > 150$ MeV are measured with a resolution of
$\sigma_{p_{\perp}}/p_{\perp} = 0.003 \oplus 0.0008 \cdot p_{\perp}$.
It was assumed that electrons with energies of at least 10 GeV and
$\theta > 30$ mrad could be identified.

First, large samples of single-tag events (approximately 12 fb$^{-1}$)
were generated at a centre-of-mass energy of $E_{\rm cm} = 184$ GeV
with \hwg\ using the GRV LO \cite{GLU-9202}, SaS1D \cite{SCH-9501} and LAC1
\cite{LAC1} structure functions.  True single-tag events were defined to be
those with a tag electron in the range $60 < \theta < 100$ mrad, which for
the GRV structure functions gives a mean $Q^2$ of 40.0 GeV$^2$ and
a cross section of 11.6 pb.  These events were processed by
the detector simulation routine in order to determine a
response matrix $R$ based on each of the three sets of structure functions.

Scatter plots of the true and observed values of $x$ are shown
in Fig.~\ref{fig:xvis_vs_xtrue} for two different ranges of
$|\cos \theta_{\rm had}|$.  In
Fig.~\ref{fig:xvis_vs_xtrue}(a), the (true) hadronic system has an angle of
at least 60$^{\circ}$ with the beam line.  This results in a relatively
good measurement of the hadronic mass $W$, and hence in a good resolution
for $x$.  In Fig.~\ref{fig:xvis_vs_xtrue}(b), the hadronic system
has $|\cos \theta_{\rm had} | > 0.9$, so that a larger fraction of
the hadrons escape detection.  This leads to a visibly weaker
correlation between the measured and true values of $x$.

\setlength{\unitlength}{1.0 cm}
\renewcommand{\baselinestretch}{0.8}
\begin{figure}[bht]
\begin{picture}(10.0,4.3)
\put(-1.3,-1.2){\includegraphics{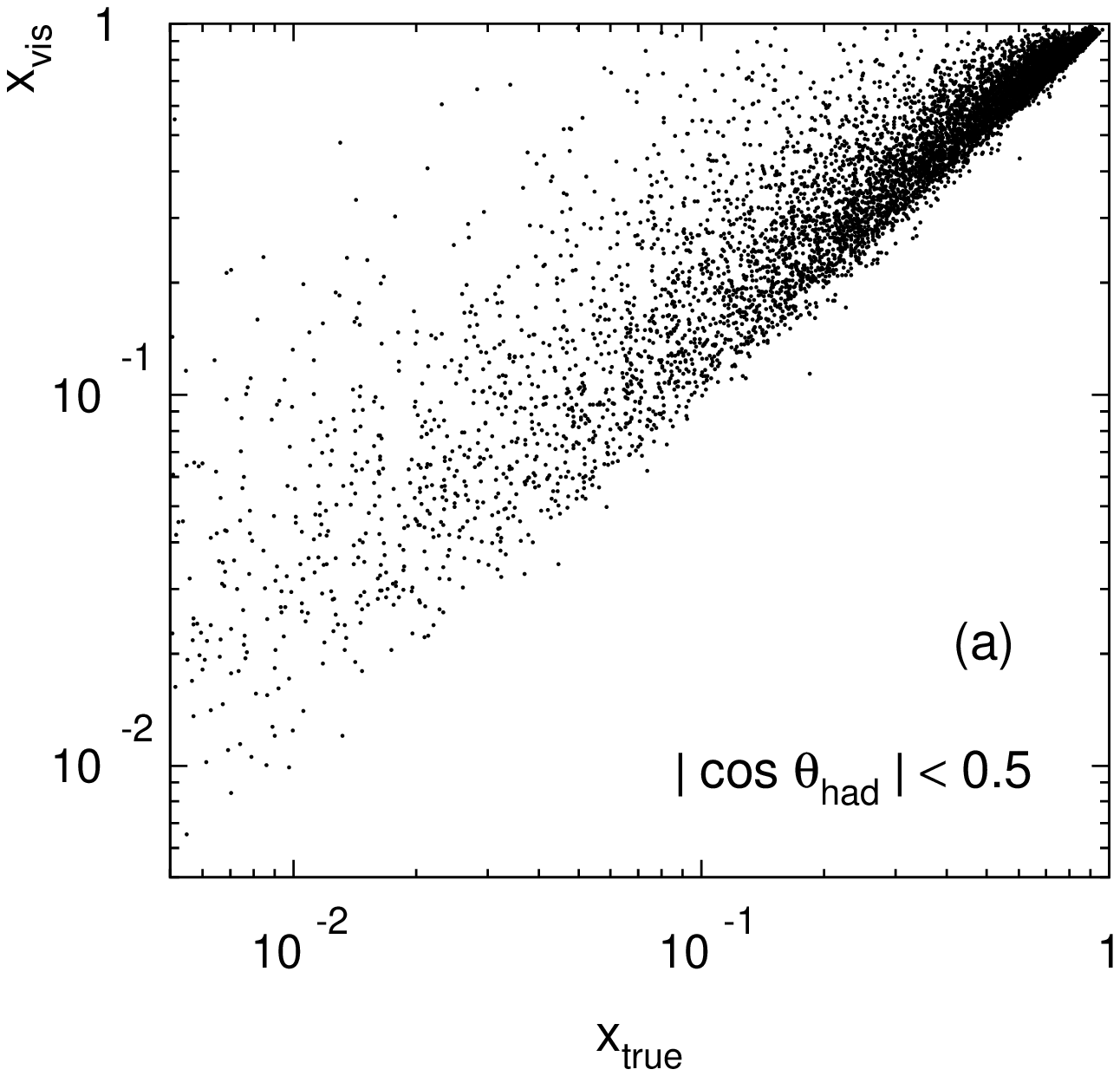}}
\put(3.6,-1.2){\includegraphics{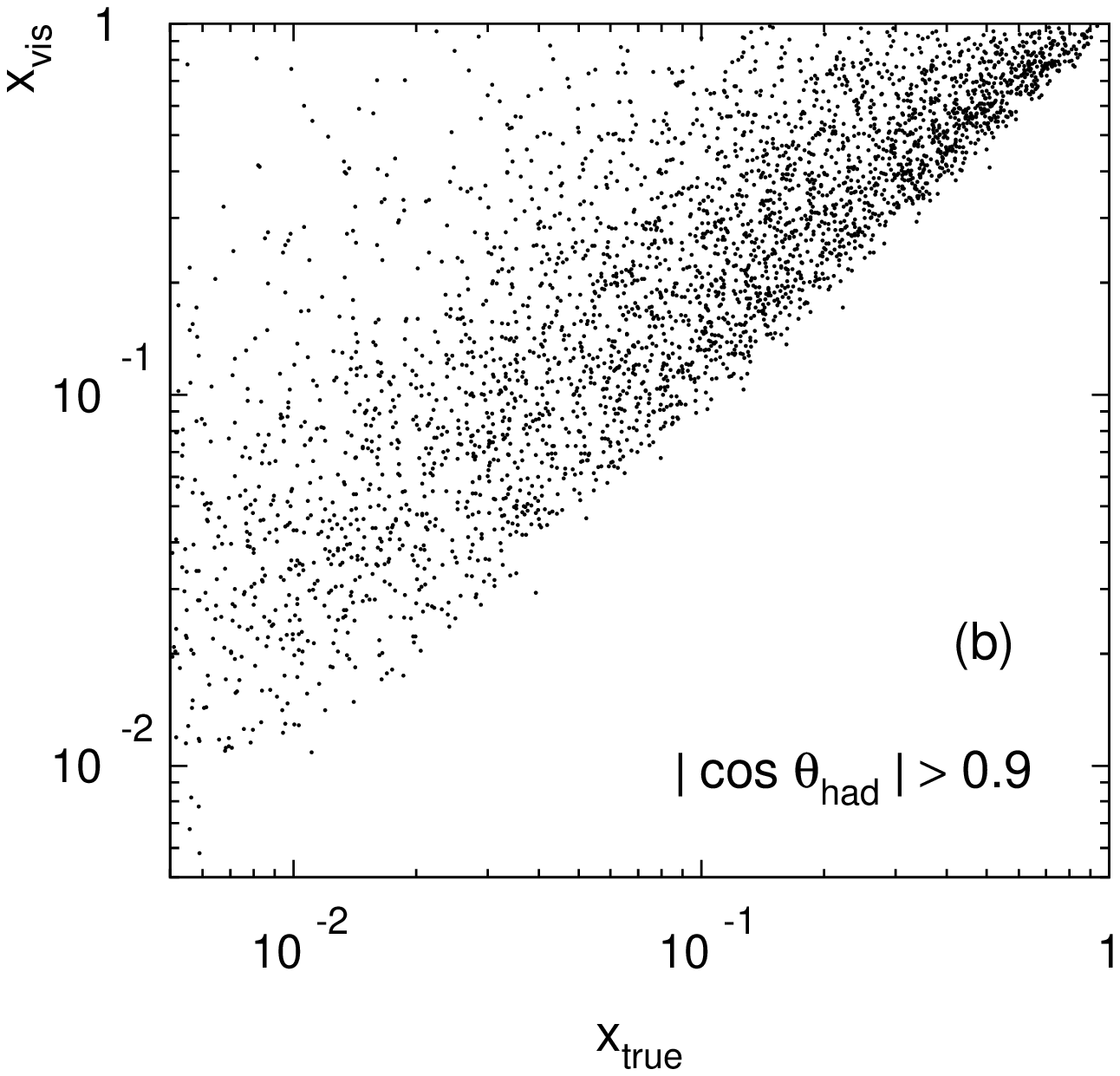}}
\put(7.9,.5){\makebox(5.2,5)[b]{\begin{minipage}[b]{5.2cm}
\caption{\footnotesize Scatter plots of the visible and true values of $x$
from the \hwg\ Monte Carlo (with GRV structure function)
and a simplified detector simulation, for two different ranges of the
angle of the hadronic system with respect to the beam line.}
\label{fig:xvis_vs_xtrue}
\end{minipage}}}
\end{picture}
\end{figure}
\renewcommand{\baselinestretch}{1}
\small\normalsize

The effect of the direction of the hadronic system on the $x$
resolution can also be seen in Fig.~\ref{fig:mean_sigma_xvis}, showing
(a) the mean and (b) the standard deviation $\sigma_x$ of
$x_{\rm vis}$ as a function of $x_{\rm true}$ for the same ranges of
$\cos \theta_{\rm had}$ as in Fig.~\ref{fig:xvis_vs_xtrue}.
For $x$ less than around 0.3, the typical $x$ resolution is between
0.12 and 0.18 for $| \cos \theta_{\rm had} | > 0.9$, but is around
0.08 -- 0.13 for $| \cos \theta_{\rm had} | < 0.5$.  Thus if two
models differ in their distributions of $\cos \theta_{\rm had}$,
they will result in different response matrices for $x$.

\setlength{\unitlength}{1.0 cm}
\renewcommand{\baselinestretch}{0.8}
\begin{figure}[thb]
\begin{picture}(10.0,4.3)
\put(-1.3,-1.2){\includegraphics{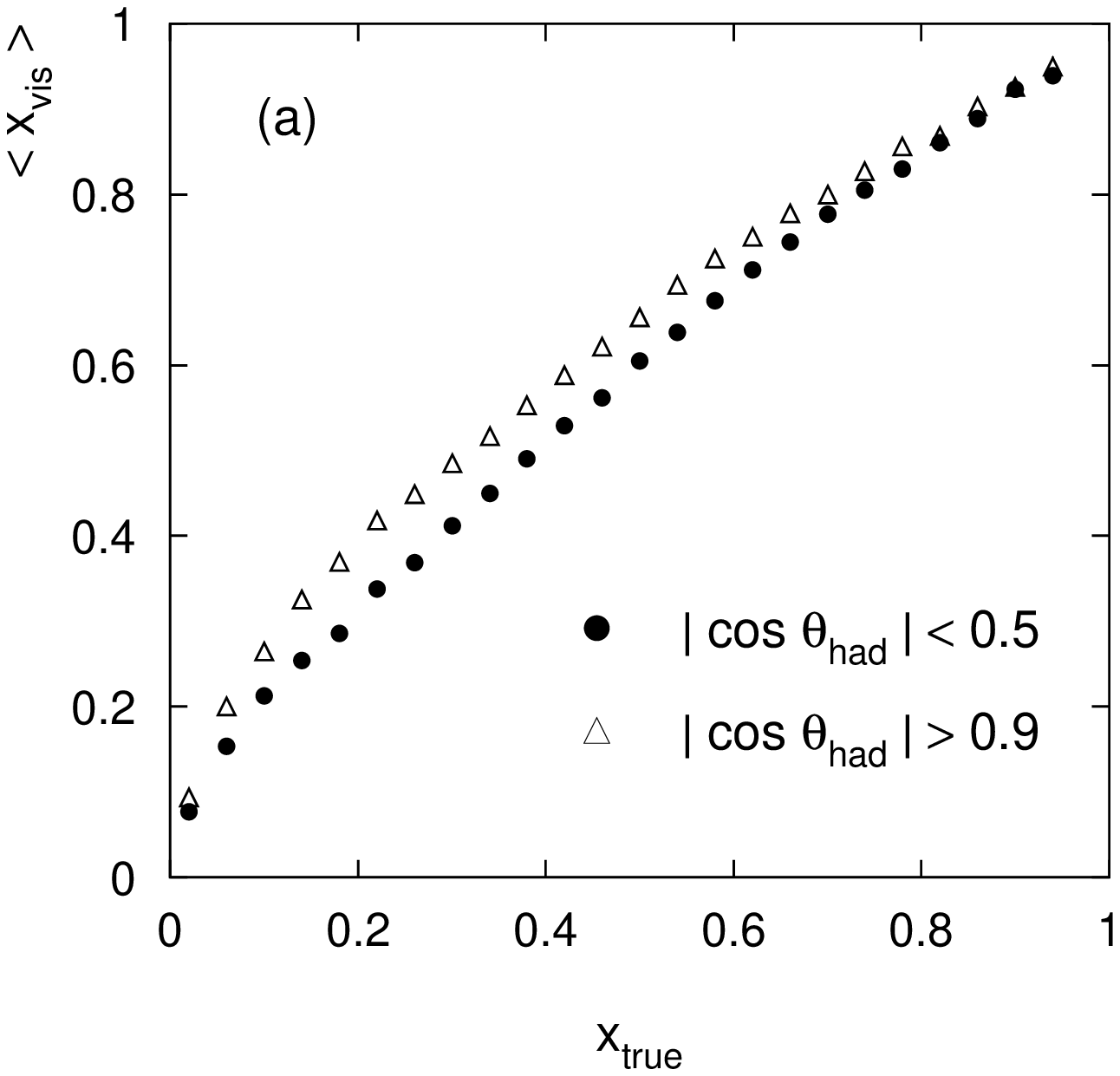}}
\put(3.6,-1.2){\includegraphics{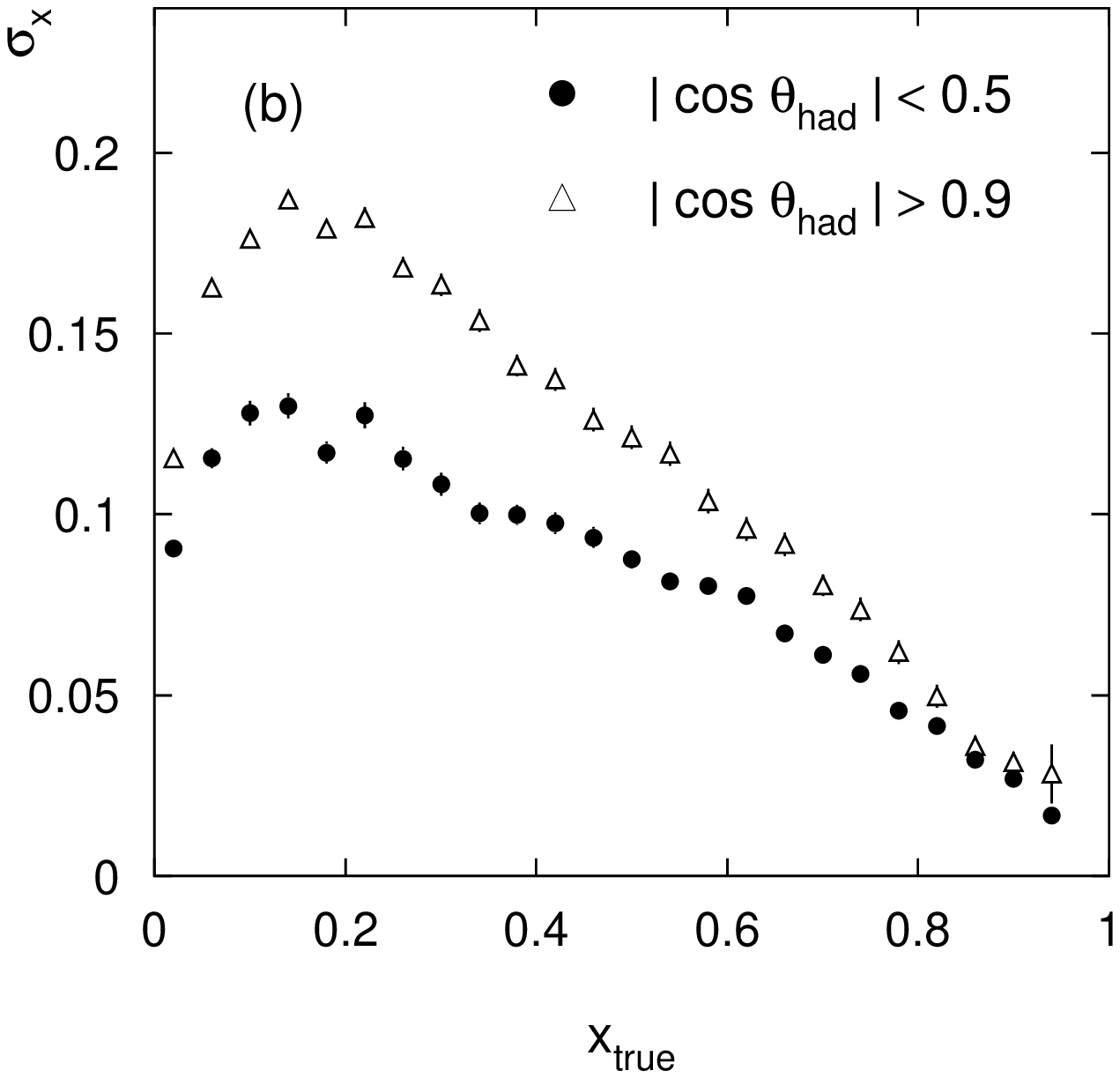}}
\put(7.9,.5){\makebox(5.2,5)[b]{\begin{minipage}[b]{5.2cm}
\caption{\footnotesize (a) The mean and (b) the standard deviation of
the measured $x$ distribution as a function of the true value of $x$,
for events in two ranges of $\cos \theta_{\rm had}$.}
\label{fig:mean_sigma_xvis}
\end{minipage}}}
\end{picture}
\end{figure}
\renewcommand{\baselinestretch}{1}
\small\normalsize

In order to minimize the errors in the unfolded result, the bin size should
not be too much smaller than the resolution.  The bins should be
sufficiently small, however, so that the resolution is approximately
constant over a bin.  At low $x$, these two considerations come into
conflict.  The bin boundaries for $x$ were chosen to be
$[0.0, 0.05, 0.1, 0.2, 0.3, 0.4, 0.6, 0.8, 1.0]$ (8 bins).
The resolution for $|\cos \theta_{\rm had}|$ is around
0.20 -- 0.25 for $|\cos \theta_{\rm had}| < 0.5$, and improves to
around 0.18 as $|\cos \theta_{\rm had}|$ increases.  The bins for
$|\cos \theta_{\rm had}|$ were chosen to be
$[0.0, 0.3, 0.5, 0.7, 0.8, 0.9, 1.0]$ (6 bins).

To test the unfolding procedure, a smaller independent sample
of events was generated using the GRV structure functions and
processed by the detector simulation routine; the resulting events
were treated as real data.  The test sample corresponds to 207 pb$^{-1}$,
and consists of 2444 true single-tag events (according to the definition
above).  Of these, 1697 passed relatively loose event selection criteria,
having a tag electron measured in the range $60 < \theta < 100$ mrad, no
electron on the opposite side above 30 mrad, and in addition, at least three
charged particle tracks.

Figure~\ref{fig:dndx}(a) shows the distribution $dN/dx$ based on
large samples of data generated with the GRV, SaS1D and LAC1 structure
functions, normalized to the expected number of entries for an integrated
luminosity of 207 pb$^{-1}$.
Figure~\ref{fig:dndx}(b) shows the GRV predicted distribution of
$x_{\rm true}$, $dN_i/dx = \mu_i/\Delta x_i$, where $\Delta x_i$ is the
width of bin $i$.  Also shown are the expectation values of what
one would observe including the effects of detector acceptance and resolution,
corresponding to the histogram $\vec{\nu}$.
The histogram of $x_{\rm vis}$ from the simulated data, i.e.\ the
vector $\vec{n}$ divided by the bin widths, is shown as points with
error bars. These are subject to both detector effects and statistical
fluctuations.

\setlength{\unitlength}{1.0 cm}
\renewcommand{\baselinestretch}{0.8}
\begin{figure}[thb]
\begin{picture}(10.0,4.3)
\put(-1.4,-1.2){\includegraphics{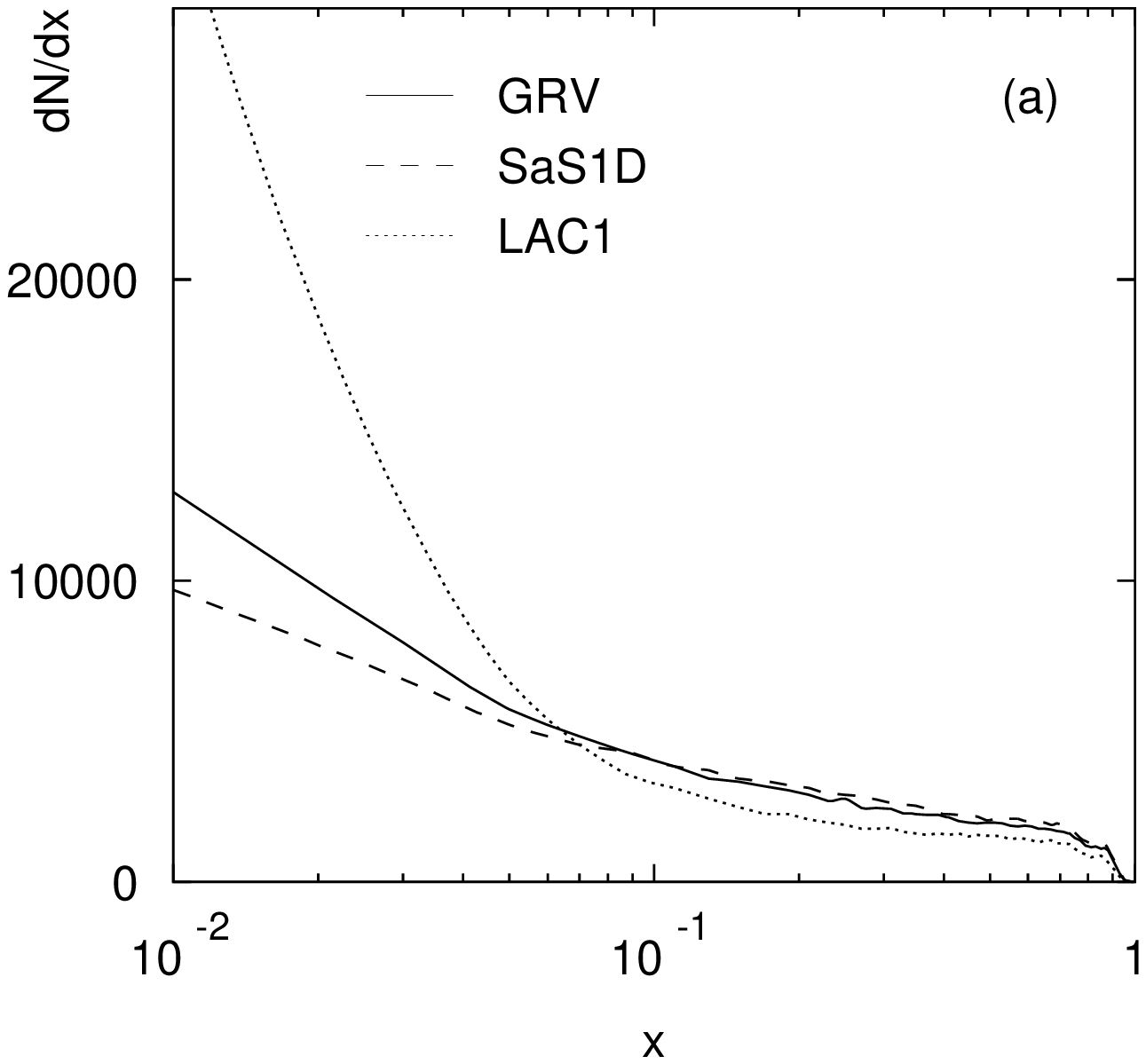}}
\put(3.6,-1.2){\includegraphics{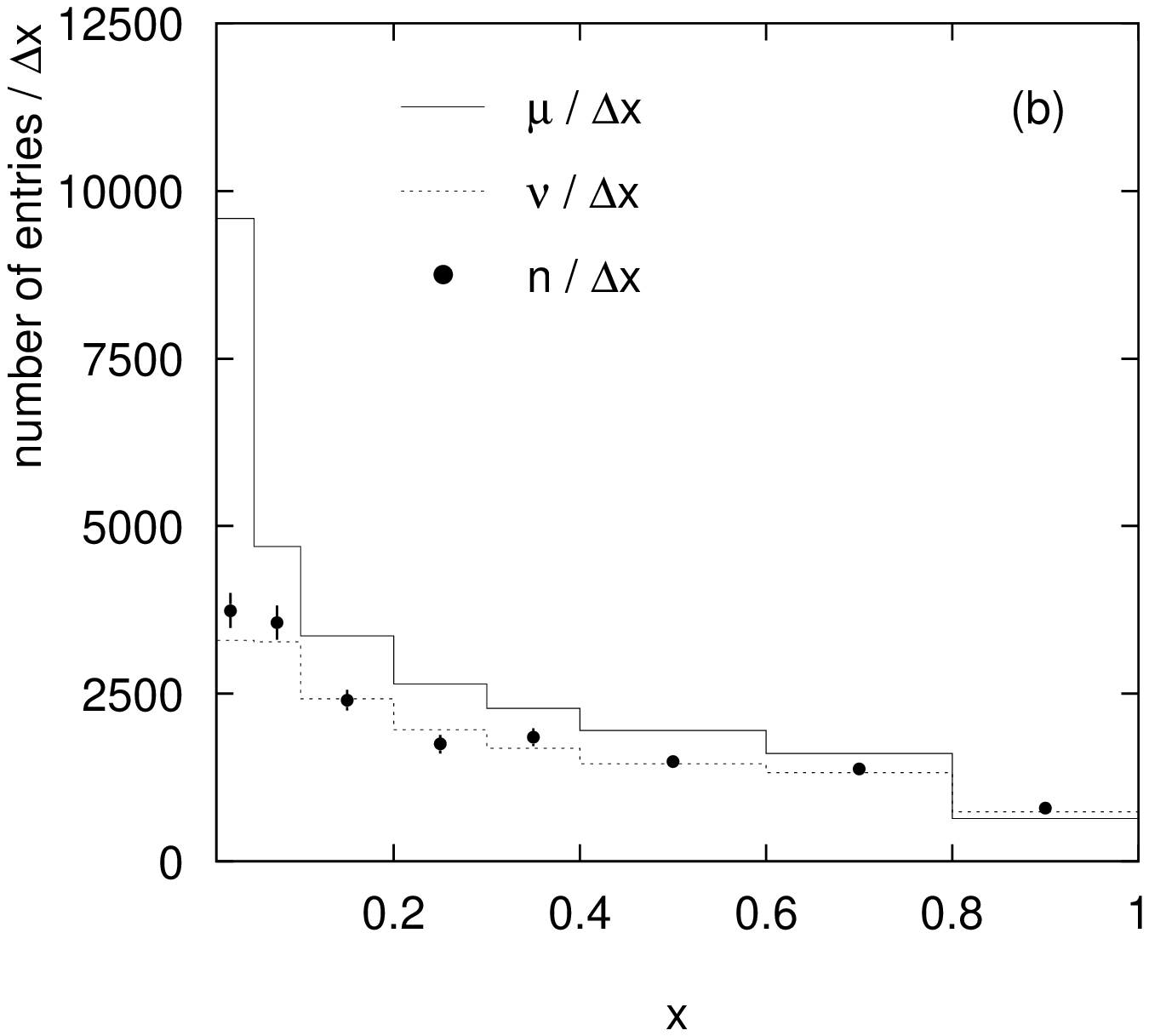}}
\put(7.9,.5){\makebox(5.2,5)[b]{\begin{minipage}[b]{5.2cm}
\caption{\footnotesize (a) Dis\-tri\-bu\-tions~$dN/dx$
from~dif\-fer\-ent struc\-ture func\-tions.  (b)
The ex\-pec\-ted $dN/dx$ from HER\-WIG with GRV (solid his\-to\-gram), the
ex\-pec\-ta\-tion values af\-ter de\-tec\-tor smear\-ing (dotted
his\-to\-gram) and the ob\-served num\-ber of
en\-tries from the test data sample (points).}
\label{fig:dndx}
\end{minipage}}}
\end{picture}
\end{figure}
\renewcommand{\baselinestretch}{1}
\small\normalsize

First, the difference between Tikhonov and MaxEnt regularization
was investigated using the test data sample and response matrix
both based on GRV.  Figure \ref{fig:dndx_tiko_maxent} shows the result of a
one-dimensional unfolding of the $x$ distribution, with statistical error
bars, using (a) Tikhonov and (b) entropy-based regularization functions.  The
regularization parameter $\alpha$ was determined in both cases such that
the biases of the estimators, $b_i = E[\hat{\mu}_i] - \mu_i$,
are consistent with zero within their own statistical errors.  This
is done by constructing estimators $\hat{b}_i$ for the biases;
cf.\ \cite{glenstat}.  Here this criterion leads to larger
errors for the MaxEnt case.  Because of this fact, however, MaxEnt gives
a better level of agreement between the estimated and true distributions
within the statistical errors.  The bias in the Tikhonov result can be
reduced by increasing the regularization parameter, in which case it
becomes similar to that shown here from MaxEnt.
The prescription for setting $\alpha$ is by no means
unique, and a corresponding systematic uncertainty should be assigned
to the final result.  A reasonable measure of this uncertainty is given
by the estimated bias.

\setlength{\unitlength}{1.0 cm}
\renewcommand{\baselinestretch}{0.8}
\begin{figure}[thb]
\begin{picture}(10.0,4.3)
\put(-1.4,-1.2){\includegraphics{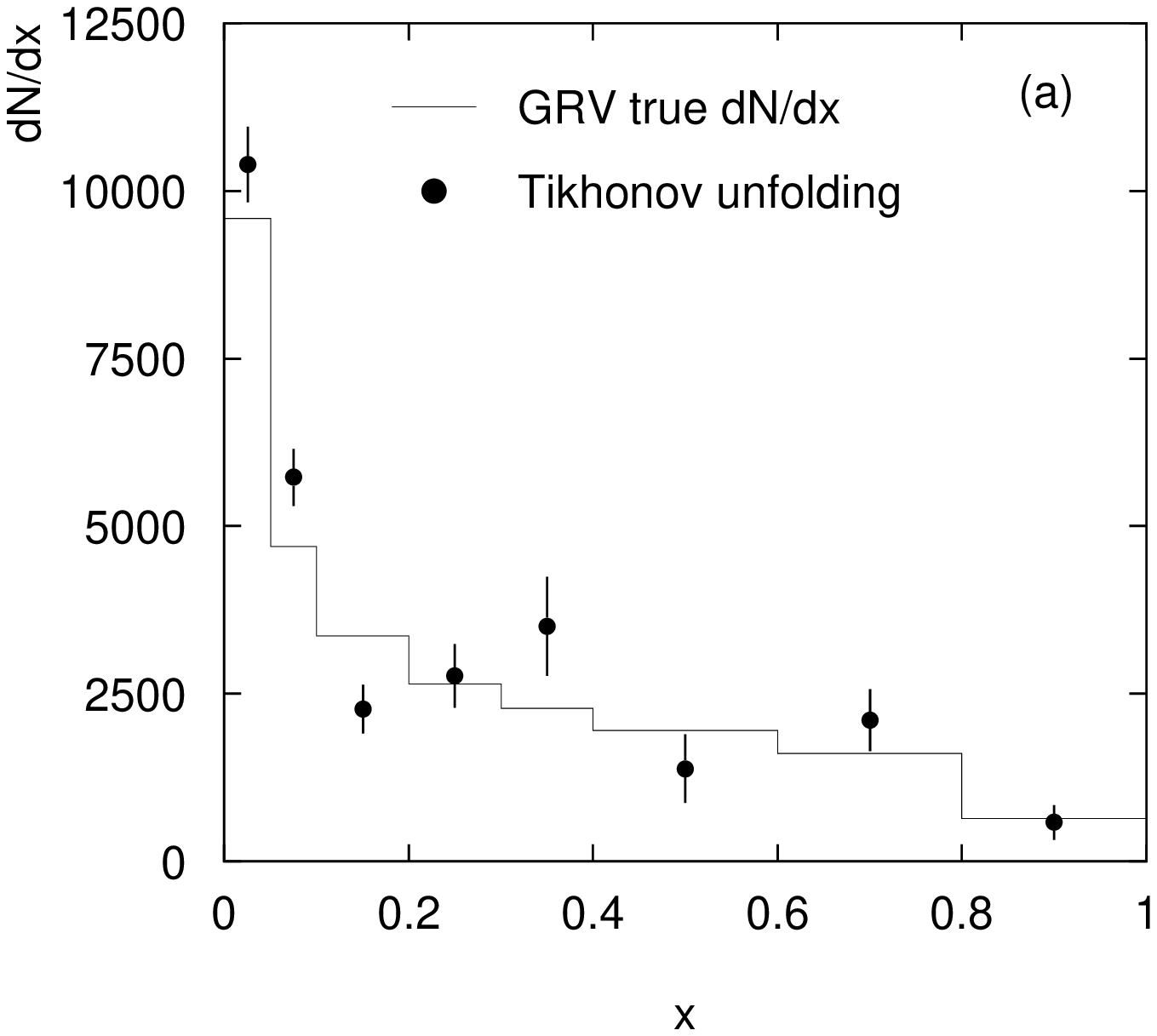}}
\put(3.6,-1.2){\includegraphics{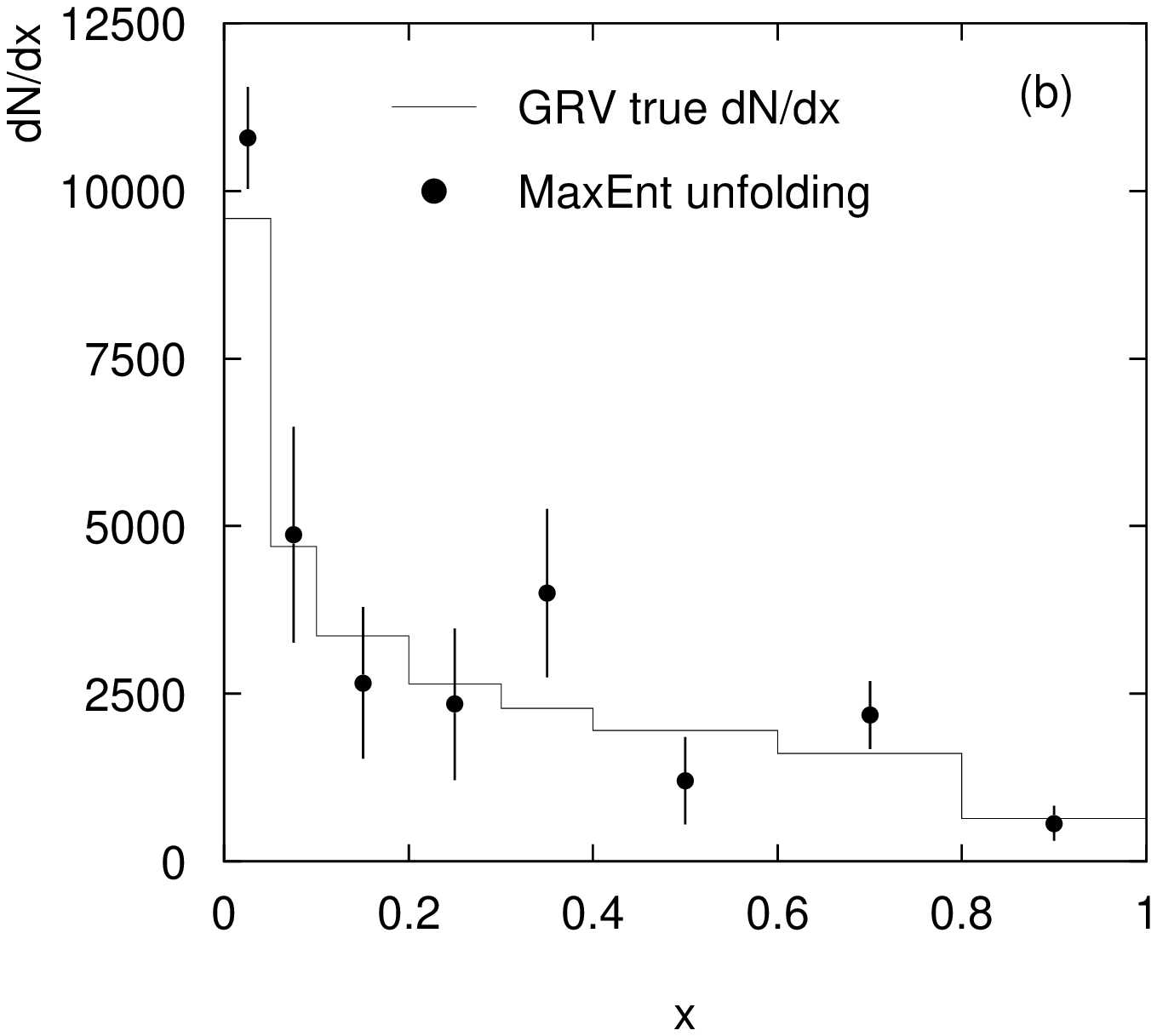}}
\put(7.9,.5){\makebox(5.2,5)[b]{\begin{minipage}[b]{5.2cm}
\caption{\footnotesize Unfolded distributions $dN/dx$
using (a) Tikhonov and (b) entropy-based regularization functions.}
\label{fig:dndx_tiko_maxent}
\end{minipage}}}
\end{picture}
\end{figure}
\renewcommand{\baselinestretch}{1}
\small\normalsize

Next, the question of model dependence of the response matrix was
investigated using MaxEnt regularization.  Figure~\ref{fig:dndx_1d}(a)
shows results obtained using response matrices from \hwg\ with GRV,
SaS1D and LAC1 structure functions, where the test data sample was
in all three cases based on GRV.  All of the results are in reasonable
agreement with the true values within their statistical errors.  From the
ratio of unfolded to true distributions in Fig.~\ref{fig:dndx_1d}(b),
one can see that the unfolded result indeed depends on which
response matrix is used.

\setlength{\unitlength}{1.0 cm}
\renewcommand{\baselinestretch}{0.8}
\begin{figure}[thb]
\begin{picture}(10.0,4.3)
\put(-1.1,-1.2){\includegraphics{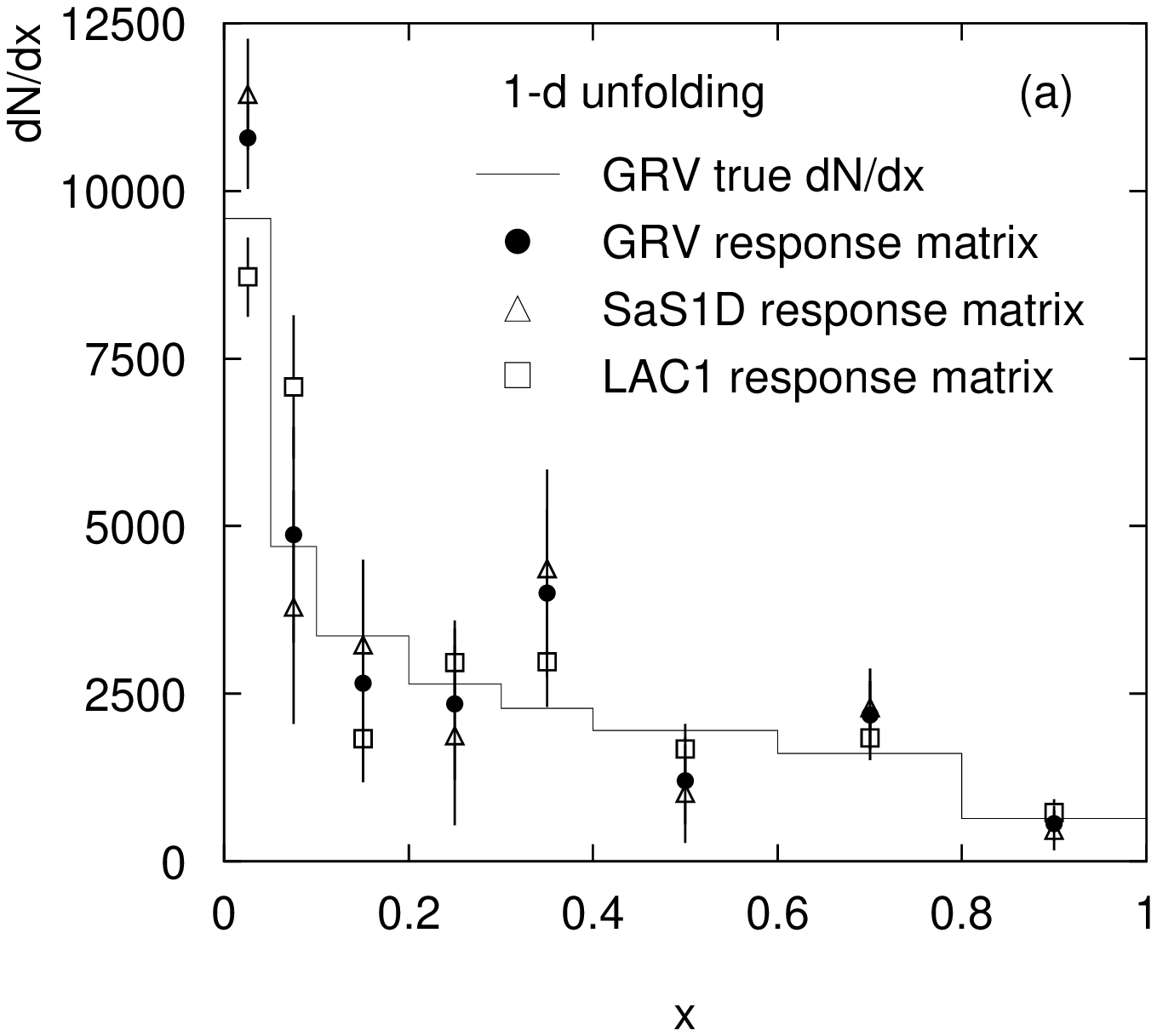}}
\put(3.6,-1.2){\includegraphics{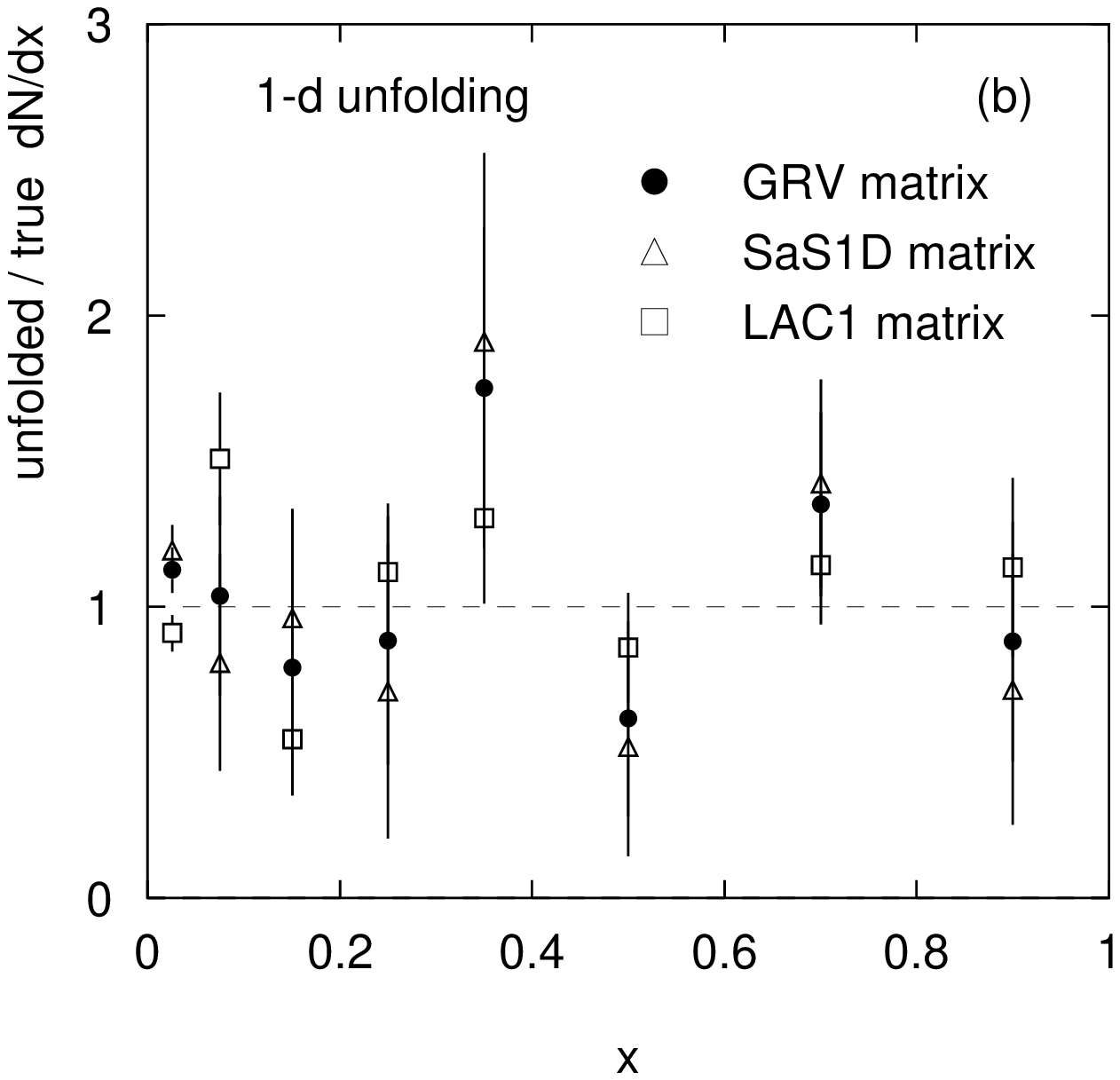}}
\put(7.9,.5){\makebox(5.2,5)[b]{\begin{minipage}[b]{5.2cm}
\caption{\footnotesize Results from a
one-di\-men\-sion\-al~un\-folding of the var\-i\-able $x$.  (a)
The true and un\-folded dis\-tri\-bu\-tions and (b) their ra\-tios.}
\label{fig:dndx_1d}
\end{minipage}}}
\end{picture}
\end{figure}
\renewcommand{\baselinestretch}{1}
\small\normalsize

Figure \ref{fig:dndx_2d} shows the result of a two-dimensional unfolding
of the variables $x$ and $|\cos \theta_{\rm had}|$, again
using the entropy-based regularization function, and the same criterion
as before for determining the regularization parameter.
Again, all of the results are in reasonable agreement with the true values
within the statistical errors.  Here, however, the spread between
the points for different models is reduced with respect to what was
obtained from the one-dimensional unfolding.  The differences between
results from GRV and SaS1D matrices differ by 1 -- 5\%,
except for a 9\% change for the second bin; these shifts are much smaller
than the relative statistical errors of the unfolded $\hat{\mu}_i$.

\setlength{\unitlength}{1.0 cm}
\renewcommand{\baselinestretch}{0.8}
\begin{figure}[thb]
\begin{picture}(10.0,4.3)
\put(-1.1,-1.2){\includegraphics{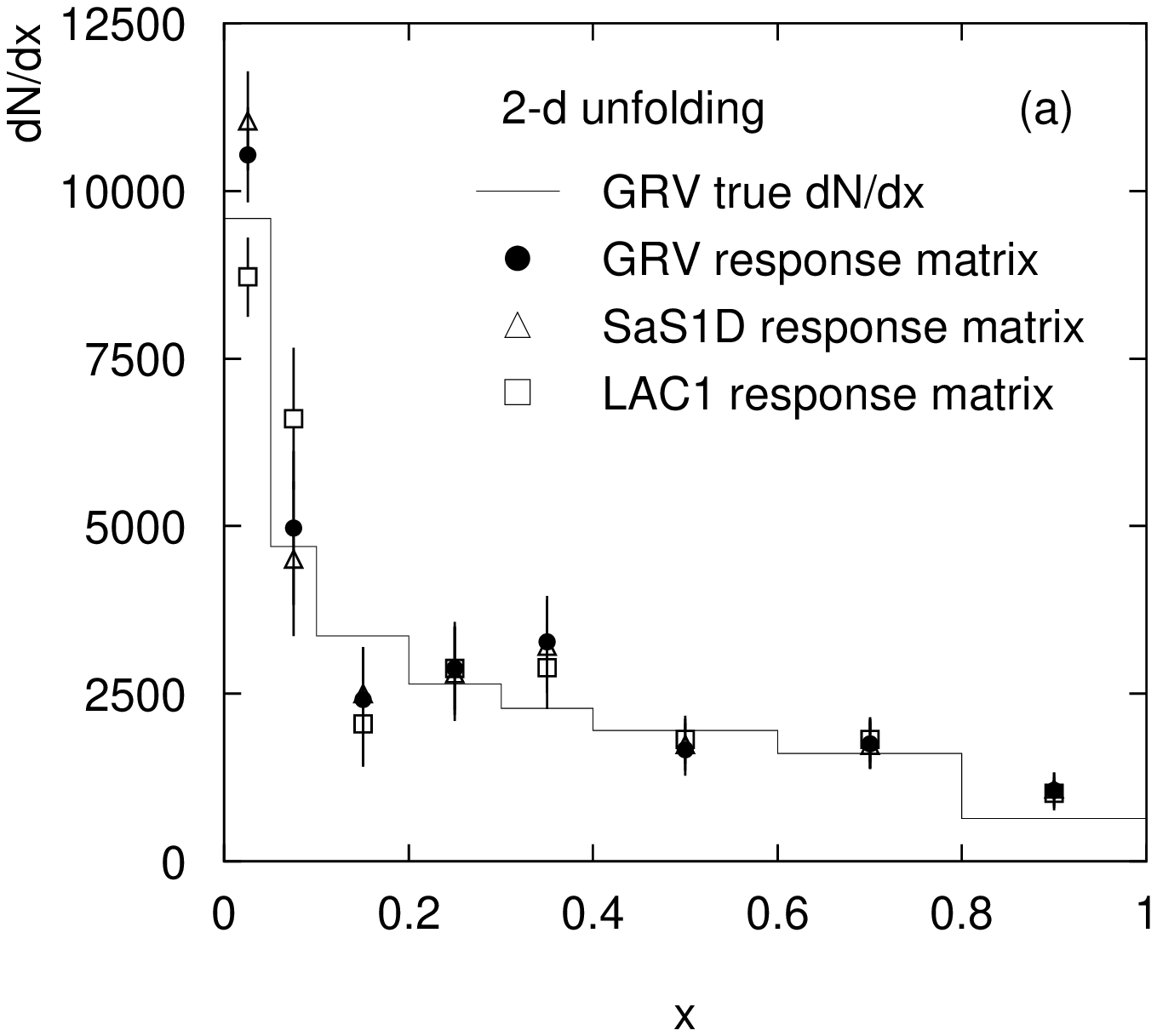}}
\put(3.6,-1.2){\includegraphics{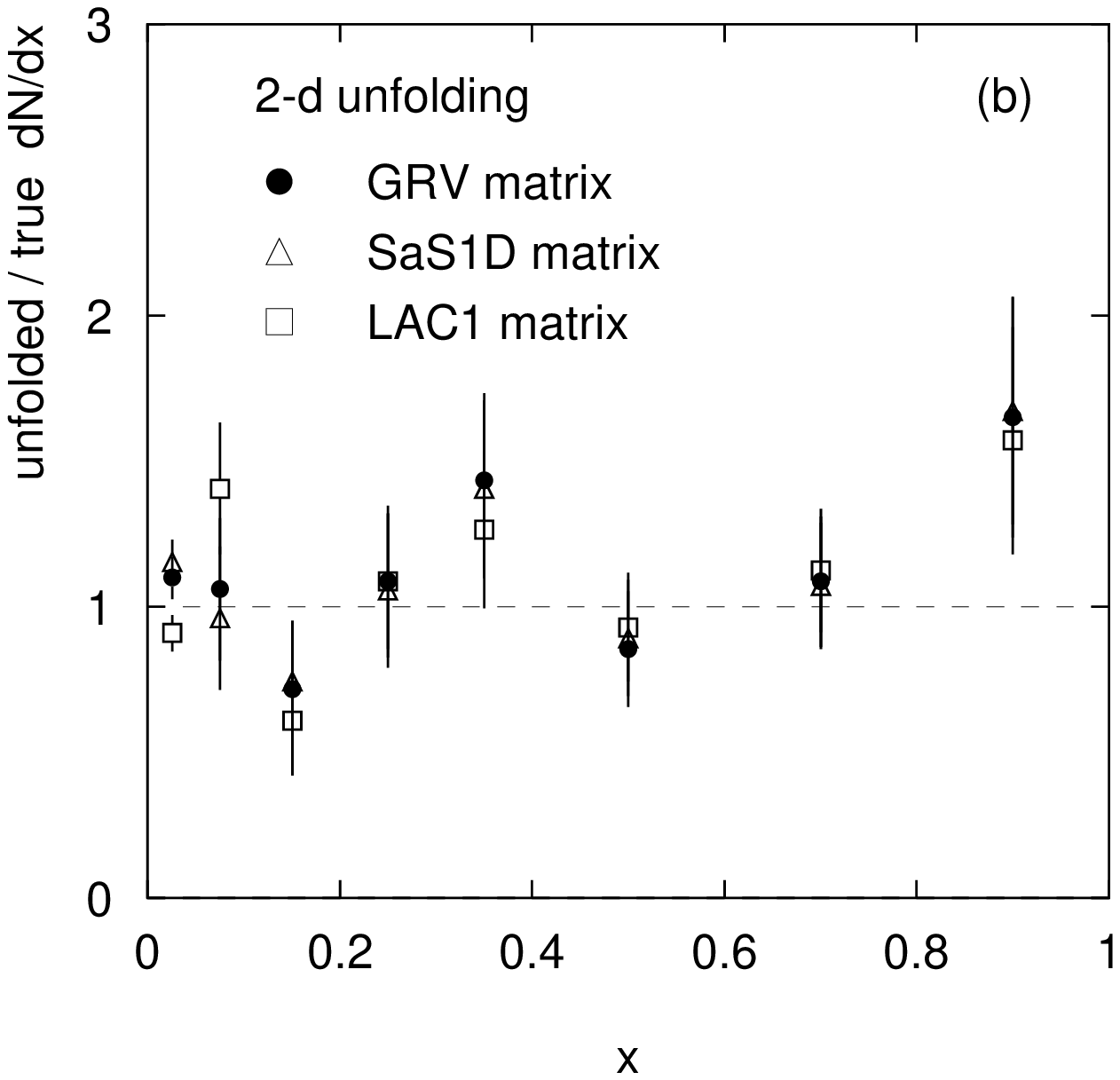}}
\put(7.9,.5){\makebox(5.2,5)[b]{\begin{minipage}[b]{5.2cm}
\caption{\footnotesize Results of a
two-di\-men\-sion\-al unfold\-ing of the var\-i\-ables $x$ and
$\cos \theta_{\rm had}$.  (a) The true and un\-fold\-ed dis\-tri\-bu\-tions
and (b) their ra\-ti\-os.}
\label{fig:dndx_2d}
\end{minipage}}}
\end{picture}
\end{figure}
\renewcommand{\baselinestretch}{1}
\small\normalsize

It is not entirely clear why the entropy-based regularization
gives smaller statistical errors in the two-dimensional case than when
unfolding in one dimension.  Although the variable $\cos \theta_{\rm had}$
should reduce the model dependence of the response matrix, $x$
and $\cos \theta_{\rm had}$ are only weakly correlated, and hence
the angle of the hadronic system does not directly provide information about
$x$.  It appears here that the prescription for determining
the regularization parameter led to a solution with smaller errors
in the two-dimensional case.  It is not obvious whether this
is particular to the example done here, or rather is true more generally.

Although some improvement with two-dimensional unfolding was anticipated,
the exact reasons for the differences between the one and two-dimensional
cases here are not yet fully understood.  A large reduction in model
dependence would be expected if the models had very different distributions
of $\cos \theta_{\rm had}$.  This distribution does not change much,
however, when the \hwg\ model is used with different structure functions,
and hence the three models considered here are not very different in
this regard.  Further investigation with other models in which the
$\cos \theta_{\rm had}$ distributions show greater differences will
indicate whether the two-dimensional procedure in fact gives a significant
improvement in the result.

\section{Exclusive final states}

In this section we discuss exclusive reactions of the type
\begin{equation}
  \gamma\gamma \to n\mbox{~hadrons},
\end{equation}
where $n$ is small, 2, 3 or 4.  In particular, we focus on the case of
resonant production of a single hadronic state, $X$,
\begin{equation}
  \gamma\gamma \to X \to n\mbox{~hadrons}.
\end{equation}
This provides a direct measurement of the two-photon width of the state
$X$, $\Gamma(X\to\gamma\gamma)$, which is an important input to the
understanding of meson spectroscopy, as we discuss in
section~\ref{exclthry}.  In section~\ref{exclexpt} we discuss the extent
to which such measurements might be made at LEP2.

\subsection{Motivation}
\label{exclthry}

Two photon collisions provide a sensitive flavour-dependent probe of the
$\mathrm{q\bar{q}}$ spectrum.  It is important to establish the
spectroscopy of the 1--2.5~GeV region, where many of the assignments
that have been made are only tentative.  One of the most important
questions in the strongly-interacting limit of QCD is whether or not
purely gluonic bound states exist.

In recent years lattice QCD calculations have improved enormously in
precision.  The systematic errors are under good control, and several
different calculations come to the same conclusion:  that the lowest
lying gluonic state has $J^{PC}=0^{++}$ and a mass of
\begin{equation}
  m_G = 1.61\pm0.07\pm0.13\;\mathrm{GeV},
\end{equation}
where the first error is statistical and the second is a systematic
error estimated from the remaining discrepancies between different
calculations\cite{FEC1}.  There is also evidence that the $2^{++}$ and
$0^{-+}$ states lie around 2~GeV, although here the calculations have
considerably larger uncertainties.  Lattice calculations also give us
the width of the $0^{++}$ decay to pseudoscalars,
\begin{equation}
  \Gamma_G \sim 100\;\mathrm{MeV},
\end{equation}
although again with considerable uncertainty.

Not surprisingly, one expects glueballs to be produced most easily in
gluon-rich environments.  The classic example is of $J/\psi$ radiative
decays, in which the lowest-order perturbative contribution is
\begin{equation}
  J/\psi \to \gamma\,g\,g,
\end{equation}
where the gluon pair must be in a colour-singlet state, since both the
$J/\psi$ and photon are colour singlets.  Thus this is ready to be
projected directly on to a glueball wave function.  Comparing this with
the production of a normal meson, in which the gluons must first
annihilate to a $q\bar q$ pair, which are then projected on to the meson
wave function, we can estimate
\begin{equation}
  \frac{\Gamma(J/\psi\to\gamma G)}{\Gamma(J/\psi\to\gamma M)}
  \sim \frac1{\alpha_s^2}.
\end{equation}

The exact opposite is true in two photon collisions, which are
quark-rich environments.  A glueball can only be produced via the
annihilation of a $q\bar q$ pair into a pair of gluons, whereas a normal
meson can be produced directly, so we can estimate
\begin{equation}
  \frac{\sigma(\gamma\gamma\to M)}{\sigma(\gamma\gamma\to G)}
  = \frac{\Gamma(M\to\gamma\gamma)}{\Gamma(G\to\gamma\gamma)}
  \sim \frac1{\alpha_s^2}.
\end{equation}
The {\em stickiness} of a mesonic state is defined as
\begin{equation}
  S_X = \frac{\Gamma(J/\psi\to\gamma X)}{\Gamma(X\to\gamma\gamma)}.
\end{equation}
We expect the stickiness of all normal mesons to be comparable, while
for glueballs, we expect it to be enhanced, from the above arguments, by
a factor
\begin{equation}
  \frac{S_G}{S_M} \sim \frac1{\alpha_s^4}.
\end{equation}
Bearing in mind that this $\alpha_s$ refers to a very low
momentum-transfer process, we should estimate $\alpha_s\sim\frac12$, and
hence $S_G/S_M\sim20$.

Measuring the two-photon cross section has always therefore been
considered a classic test of the nature of a putative glueball.

Glueball phenomenology has recently been revolutionized by the
realization that the glueball is likely to mix significantly with other
$0^{++}$ mesons.  In particular, if it has mass $\sim1600$~MeV, it
should lie in the middle of a multiplet whose light $I\!\!=\!\!0$ member
is the $f_0(1370)$ and which is expected to have a
hidden-strange member at around 1600--1700~MeV.  Because of this mixing,
many of the classic phenomenological differences between glueballs and
normal mesons will be reduced.  For example the stickiness of a mixed
glueball-normal meson should be between those of pure glueball and pure
non-glueball states.  Since the wave function of a mixed state can be
estimated as
\begin{equation}
  \label{glumix}
  |G\rangle = |G_0\rangle +
  \sum |Q\rangle \frac{\langle Q|H_I|G \rangle}{E_Q-E_G},
\end{equation}
the precise properties of the mixed states are strongly dependent on
where in the multiplet the gluonic state lies, i.e.~on the energy
difference $E_Q\!-\!E_G$.

The Crystal Barrel experiment have observed two mesons with unusual
properties in exactly the region one expects mixed glueball and hidden
strange states:
\begin{equation}
  \begin{array}{ccc}
    J= & 0    & 0\mbox{~or~}2, \\
    m= & 1.50 & 1.71, \\
    \Gamma= & 120\pm20 & 140\pm12.
  \end{array}
\end{equation}
In particular, attention has focused on the 1500 state as the most
promising glueball candidate we have.  It has unusual decays, being
about 50\% to $4\pi$, apparently more through $\sigma\sigma$ than
$\rho\rho$, and with the $\eta\eta$ and even $\eta\eta'$ channels
dominating the $\mathrm{K\overline{K}}$ channel.  This pattern of decays
can be understood as a result of the mixing in
Eq.~(\ref{glumix}),
\begin{equation}
  |G\rangle \sim |G_0\rangle +
  {\cal{O}}(H_I)\left(a|\mathrm{n\bar{n}}\rangle -
    b|\mathrm{s\bar{s}}\rangle\right),
\end{equation}
where $\mathrm{n\bar{n}} = \frac1{\surd2} (\mathrm{u\bar{u}} +
\mathrm{d\bar{d}})$, and $a,b$ are positive numbers.  The destructive
interference between the light and strange components suppresses the
decay to kaons.

Similarly, the two photon width of the state should be strongly
dependent on the mass differences between the different states.
Therefore it is crucial in order to understand the gluonic sector of the
meson spectrum to measure $\Gamma_{\gamma\gamma}$ for the 1370, 1710 and
especially the 1500 states, or to place limits on them if they are not
observed.  If the 1500 state really is a glueball-dominated meson, one
expects $\Gamma_{\gamma\gamma}\sim\mathrm{few}\times0.1$~keV at most,
with a possibility that it could be as small as 0.03~keV\cite{FEC2},
so the challenge for LEP2 is to measure it to this accuracy, or to at
least place limits significantly better than 1~keV.

Another gluon-rich environment where glueballs should be abundantly
produced is in central hadron collisions,
\begin{equation}
  pp \to ppX.
\end{equation}
Indeed this is precisely where the Crystal Barrel experiment see the
1500 candidate.  It has been observed by the WA102 experiment\cite{FEC3}
that the kinematics of the scattered protons significantly affects the
production of glueball candidates, particular the dependence on their
relative azimuth\cite{FEC4}.  If both protons come out on the same side,
giving the central system a relatively high transverse momentum, the
glueball candidates are enhanced relative to the known normal mesons.
If they come out on opposite sides, they roughly balance in transverse
momentum and the central system is given little $p_t$ kick.  The
glueball candidates are then suppressed at the expense of the normal
hadrons and the continuum background.  While this is clearly an
intriguing effect, it is not well understood, and two photon collisions
might shed considerable light on this, if it is possible to isolate a
glueball signal in the double-tagged channel,
\begin{equation}
  e^+e^- \to e^+e^-X,
\end{equation}
with both the electrons detected at very small angles.  If a positive
signal can be isolated in this channel, one can then play the same
kinematic games as in central production, to try to understand the
glueball filter mechanism in more detail.

We finish this section by briefly mentioning hybrid mesons.  These are
high-mass states of normal mesons, whose gluonic degrees of freedom have
been excited.  Excited pion states have been observed at 1300 and
1800~MeV in fixed-target pion scattering,
\begin{equation}
  \pi N \to \pi^* \to 3\pi,
\end{equation}
but have never been observed in two photon collisions.  In pion
scattering, since the original pion is charged, the three pions can all
be charged, $\pi^\pm \to \pi^\pm\pi^+\pi^-$ or $\pi^\pm\pi^0\pi^0$, but
in two photon collisions, at least one must be neutral,
\begin{equation}
  \gamma\gamma \to \pi^0\pi^+\pi^- \mbox{~or~} \pi^0\pi^0\pi^0,
\end{equation}
making this an extremely difficult channel to trigger on.

To summarize, there is strong evidence that there are mixed
glueball-nonglueball states at 1500 and 1710, with the former being
predominantly gluonic.  Our understanding of this area of hadron physics
could be significantly improved if the LEP experiments can measure the
two photon width of the glueball candidates, or set limits of
significantly better than 1~keV on them.  Around 50\% of the decays are
to four charged pions.  If a signal can be isolated in the double-tagged
channel, considerable light could be shed on the glueball production
mechanism, by studying the dependence on the electron kinematics.

\subsection{Feasibility at LEP2}
\label{exclexpt}

As reported above, the $f_0(1500)$ resonance, observed by the Crystal
Barrel collaboration in $\mathrm{p\bar{p}}$ interactions at rest
\cite{barrel}, is a strong candidate to be the lightest scalar glueball,
lining up well with lattice QCD predictions.  Information on production
of this resonance in two-photon collisions at LEP is important in
establishing its nature, whether a pure glueball or a quark/gluon
mixture due to mixing with a $\mathrm{q\bar{q}}$ nonet.

Preliminary results are now available from the ALEPH collaboration
\cite{me} that suggest the $f_0(1500)$ does indeed have a suppressed
two-photon width and hence could be consistent with a glueball
interpretation.  Although the dominant decay mode of the $f_0(1500)$ is
to four pions, the acceptance for this final state in $\gamma \gamma$
collisions at LEP turns on just below the 1500 \mev\ mark, so a limit on
$f_0(1500)$ production in this channel is difficult to establish.
Instead the two pion final state ($\sim 20\%$ of decays) is used, where
\dedx\ information for the charged tracks is used to identify the
particles.

\begin{figure}
\epsfxsize=8cm
\epsfysize=8cm
\begin{center}\mbox{\epsfbox{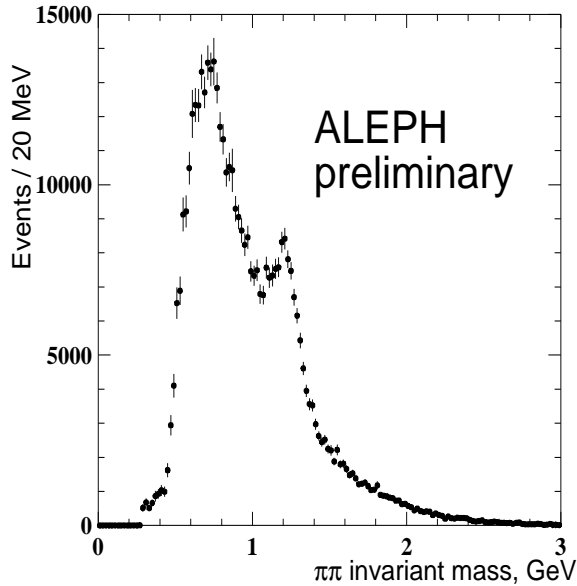}}\end{center}
\caption{\label{piccie}
The invariant mass distribution for two-pion final states.}
\end{figure}

The $\pi^+ \pi^-$ invariant mass distribution, for 160.9 ${\mathrm{pb}^{-1}}$
of ALEPH data taken from 1990 to 1995, is shown in figure \ref{piccie}.
The mass resolution is about 10~\mev, significantly  smaller than the
bin width.
A clear peak in the spectrum is seen just above 1~\gev: this can
be identified with the known tensor resonance $f_{2}(1270)$. Background
is attributed to misidentified, low-energy muon pairs from
the process $\gamma \gamma \rightarrow \mu \mu$, and also to the
$\pi\pi$ continuum.

A reasonable fit ($\chi^2$/d.o.f.=1.1) can be made to the data between
0.8 and 2.5 \gev, using a Breit-Wigner to describe the $f_{2}(1270)$ and
a polynomial for the background (although this is somewhat na\"\ii ve,
and it would be preferable to fit the detailed physics of the spectrum,
taking proper account of interference effects; however, the fit is
certainly adequate).  A window in mass from 1.38 to 1.62 \gev\
(equivalent to twice the total width of the $f_0(1500)$) is excluded
from the fit~-- this is the ``signal'' region.  Extrapolation of the fit
through this window is used to define the background to the $f_0(1500)$
signal. From a binned likelihood fit, introducing signal in the form of
a Breit Wigner resonance of mass 1500 \mev\ and width 120 \mev, an upper
limit on the number of signal events is established as 128 events at
95\% \CL

Incorporating the selection efficiency for $\gamma \gamma \rightarrow
f_0(1500) \rightarrow \pi^+ \pi^-$ (22\%) the trigger efficiency at this
mass (64\%) and the branching ratio for $f_0(1500) \rightarrow \pi^+
\pi^-$ (24\% \cite{zou}), this limit translates to an upper limit on the
two-photon width of the $f_0(1500)$ of
$$\Gamma_{\gamma \gamma} \: < \:\: 0.17 \: \kev, \:\:\: 95\% \:\: \CL $$
The lower limit on the stickiness \cite{sticky} of the state is then
calculated to be 13, where stickiness has been normalised such that the
$f_{2}(1270)$ $\mathrm{q\bar{q}}$ resonance has stickiness equal to~1.
This limit is higher than would be expected for a simple
$\mathrm{q\bar{q}}$ state, suggesting that the $f_0(1500)$ may indeed be
a glueball-dominated state.

Further study of $f_0(1500)$ production in $\gamma \gamma$ collisions
would be very useful, as statistics from LEP2 accumulate.  It seems
reasonable to hope that the limit could be improved by up to an order of
magnitude, probing the entire range expected theoretically.  Other decay
channels, such as $\mathrm{K\overline{K}}$ could also be studied.  Here
there are experimental difficulties: trigger schemes used in the LEP
experiments are not usually favourable to these low-mass processes; and
discrimination of pions and kaons by \dedx\ information is poor, but,
for example, the DELPHI experiment is equipped with ring-imaging
\v{C}erenkov devices (RICH) that should offer more effective
identification.  From comparison of the $\pi\pi$ and
$\mathrm{K\overline{K}}$ channels, the nature of the $f_0(1500)$ could
be more firmly established, and perhaps a mixing angle between the
glueball and $\mathrm{q\bar{q}}$ nonet extracted.

At present it is almost impossible to predict the precision with which
the `glueball filter' could be tested at LEP2, since ALEPH and OPAL's
low angle taggers have only recently been commissioned.  Furthermore,
since the signal has not yet been observed in untagged data, the rate of
double-tagged events must be very low indeed.  Nevertheless, if a signal
is ever observed in the double-tagged channel, testing its kinematic
dependence will clearly be a priority.

\section{Summary}

The increase in energy and luminosity from LEP1 to LEP2 offer tremendous
possibilities for two photon physics, as well as increased challenges.
Much of our experience gained from LEP1 can be directly applied to LEP2,
and in addition there have been several hardware upgrades and
improvements to the integration of detector components crucial for two
photon physics into trigger systems.  Furthermore, the backgrounds are
reduced.  However in some areas, like single-tagged deep inelastic
scattering, LEP2 represents a great leap into the unknown.  The increase
in energy allows a much greater reach into the regions of small $x$ and
large $Q^2$, which are the most theoretically interesting regions.

In two photon physics, unlike proton deep inelastic scattering at HERA
for example, the energy of the target beam is not known~-- it is spread
according to the Weizs\"acker-Williams spectrum.  Therefore the
kinematics of the scattering cannot be determined from the electron
alone, as they can at HERA, but must be determined from the hadronic
final state.  At small $x$, much of this hadronic activity is directed
towards the forward (small-angle) regions of the detector.  Roughly
speaking, the further forward one goes, the worse the hadronic energy
resolution and efficiency get, until at very small angles one loses
hadrons completely into the beam-pipe.  Therefore the measured $x$ value
depends critically on the angular distribution of hadronic energy.

We have spent much of this workshop discussing the final state of deep
inelastic scattering, particularly at small $x$.  It is now
well-established that the data are less forward-peaked than any of the
QCD-based Monte Carlo event generators.  While the theoretical
understanding of why this should be is still lacking, we have made
progress in
finding models that better describe the data.  Minor modifications to
\hwg\ and \pyth\ result in greatly improved fits, as does using the
\pho\ program for tagged collisions, despite its only being designed for
untagged collisions.

Another approach to reducing the uncertainty due to modelling of the
hadronic final state is to measure it from data.  This can be done by
unfolding in several variables simultaneously.  The first studies in
this direction were presented at this workshop, and seem very promising:
more detailed studies are clearly warranted.  In addition, the details
of how unfolding algorithms work were explored, as was the r\^ole of
different regularization procedures, which give different trade-offs
between statistical errors and bias.  In the past there has been a
tendency to treat such algorithms as `black boxes', and it is to be
hoped that this increased understanding will be reflected in improved
analyses for LEP2.

We also discussed the importance of two photon physics to meson
spectroscopy, most notably in the identification of glueball candidates.
These should be suppressed in two photon collisions relative to `normal'
mesons, but be enhanced in gluon-rich environments like central
production.  Attention has been particularly focusing on the
$f_0(1500)$ as the most likely glueball candidate.  First results from
ALEPH on its photonic width were presented, showing that it is indeed
suppressed~-- no signal was seen, and a limit of $\Gamma_{\gamma\gamma}
\bigl(f_0(1500)\bigr) < 0.17\:\kev$ was set with 95\% confidence.  Since
glueballs should mix with normal mesons with the same quantum numbers,
this width should not be too suppressed, and it is likely that with the
full statistics of LEP2, either a non-zero width will be measured, or a
limit severe enough to rule out current models obtained, significantly
improving our understanding of hadron spectroscopy.

Progress was made in all these areas at the workshop, which should have
a significant impact on the study of two photon physics at LEP2.

\ack
We would like to thank the organizers of the workshop for a stimulating
and productive meeting.  We would also like to thank the Monte Carlo
authors who could not be at the workshop, Ralf Engel, Leif L\"onnblad
and Torbj\"orn Sj\"ostrand, for patiently answering our questions
about their models.

\end{document}